\documentclass[%
jmp,
 amsmath,amssymb,
reprint,%
author-numerical,%
]{revtex4-1}
\usepackage[english]{babel}
\usepackage{graphicx}
\usepackage{subcaption}
\usepackage{dcolumn}
\usepackage{bm}
\usepackage{multirow}
\usepackage[utf8]{inputenc}
\usepackage[T1]{fontenc}
\usepackage{mathptmx}

\newcommand{\apjl}{Astrophys.~J.~Lett.}
\def\jcap{JCAP}
\def\mnras{MNRAS}
\def\aap{A\&A}

\def\physrep{Phys.~Rep.}
\def\prd{Phys.~Rev.~D}
\def\apj{Astrophys.~J.}
\def\apjs{Astrophys.~J.~Supp.}
\def\aj{Astron.~J.}

\def\baas{BAAS}

\draft

\begin{document}

\title{Prospects for detection and application of the alignment\\ of galaxies with the large-scale velocity field}

\author{I. R. van Gemeren}
\email{i.r.vangemeren@students.uu.nl}

\author{N. E. Chisari}
\email{n.e.chisari@uu.nl}
\affiliation{Institute for Theoretical Physics, Utrecht University, Princetonplein 5, 3584 CC Utrecht, The Netherlands.
}

\date{\today}

\begin{abstract}
Studies of intrinsic alignment effects mostly focus on the correlations between shapes of galaxies with each other, or with the underlying density field of the large scale structure of the universe. Lately, the correlation between shapes of galaxies and the large-scale velocity field has been proposed as an additional probe of the large scale structure. We use a Fisher forecast to make a prediction for the detectability of this velocity-shape correlation with a combination of redshifts and shapes from the 4MOST+LSST surveys, and radial velocity reconstruction from the Simons Observatory. 
The signal-to-noise ratio for the velocity-shape (dipole) correlation is 7.4, relative to 44 for the galaxy density-shape (monopole) correlation and for a maximum wavenumber of $0.2\, \mathrm{Mpc^{-1}}$. Encouraged by these predictions, we discuss two possible applications for the velocity-shape correlation. Measuring the velocity-shape correlation could improve the mitigation of selection effects induced by intrinsic alignments on galaxy clustering. We also find that velocity-shape measurements could potentially aid in determining the scale-dependence of intrinsic alignments when multiple shape measurements of the same galaxies are provided. 
\end{abstract}

\pacs{}

\maketitle 

\section{Introduction}

Intrinsic alignments are the correlations of shapes, orientations and positions of galaxies or galaxy clusters embedded in the large-scale structure in the universe. These correlations arise during formation and evolution of galaxies and clusters. An example is the deformation of the shapes of galaxies along the stretching axis of the underlying tidal field, which results in a correlation between the shape and the overall matter density field. On large scales ($> 10\,\mathrm{Mpc}$) the intrinsic alignment effects on the shapes of galaxies can be well-described by the Linear Alignment (LA) model \cite{Catelan}. On these scales Luminous Red Galaxies\label{LRG} (generally elliptical galaxies) are observed \cite{Brown, MandelbaumHirata, Joachimi, Singh, Johnston, Samuroff} to follow this model. For blue galaxies with angular momenta correlations as the relevant statistics \cite{Catelan}, there has not yet been a significant observation of alignments \cite{Mandelbaum, Johnston}. To describe intrinsic alignment effects on smaller scales, nonlinear extensions have been developed, including phenomenological ones \cite{NLAhirata, Bridle}, perturbative methods \cite{Vlah, Blazekperturb} or halo model-based approaches \cite{Schneider,Fortuna}. 

Intrinsic alignments were first studied as a contaminant to weak gravitational lensing observations \cite{Brown,GLHirata, Joachimi}. Weak lensing refers to the statistical, small and coherent deformations in the observed shapes of galaxies due to the bending of light caused by the gravitational potentials or foreground `lensing' objects. The orientations and shapes of galaxies should therefore be random to assign all the systemics that are found in these shape measurements to weak gravitational lensing. However, this is not the case in the presence of intrinsic alignment effects, so a correction is needed. Lately, intrinsic alignment effects are also studied as an interesting probe of the large scale structure which could provide cosmological information to improve, for example, the cosmological model. With statistical measurements of intrinsic alignment effects, baryon acoustic oscillations (BAOs) can be measured and primordial non-Gaussianity can be constrained better \cite{Chisari, Schmidt, Okumura1}. The former gives a standard length scale in the universe, and the latter helps determining the best fitting model for cosmological inflation \cite{Meerburg}. 

Most studies of intrinsic alignment rely on measuring the correlations between the shapes of galaxies with each other or with biased tracers of the large scale structure \cite[e.g.][]{Singh}. Recently, it has been proposed that there also exists a correlation between the shapes of galaxies and the cosmic velocity field \cite{Okumura1, OkumuraTaruya, Okumura2}. This correlation could add information on top of the already measured galaxy density-shape correlation 
because of the relation between velocity $\mathbf{v}$ and the dark matter over-density $\delta$, expressed in Fourier space as
\begin{equation}
    v(\mathbf{k}) \propto (\mathbf{k} / k^2) \delta(\mathbf{k})
\end{equation}
in linear theory. It is furthermore expected that the velocity-shape correlation signal is enlarged relative to the density-shape signal on large scales because of the $\mathbf{k}/k^2$ term, in which a small wavenumber $k$ indicates large real space distances. $N$-body simulations show that this amplification of the signal is indeed present \cite{Okumura1, OkumuraTaruya, Okumura2} and the next step would be to determine whether the velocity-shape correlation signal could be measured with real data.

In this article, the aim is to show that measuring the correlation between shapes of galaxies and the underlying velocity field of the large scale structure is within the reach of upcoming galaxy surveys. To this end, we predict this correlation relying on the linear alignment model as described in \cite{Okumura1, OkumuraTaruya, Okumura2}. Our prediction gives significant results for measuring the signal of the velocity-shape correlation in the data and therefore, we also describe two possible applications of the velocity-shape correlation. 

This article is organised as follows. First, we present a description of the statistical formalism for intrinsic alignments in section \ref{IA&stat} and use this to specify the velocity-shape correlation on large scales. This is followed in section \ref{vrec} by a description of typical reconstruction schemes of the velocity field which will be available in the near future. We perform a Fisher forecast to predict whether the signal is detectable. This is described in section \ref{FF}. After showing the results of this forecast, two possible applications for the velocity-shape correlations are mentioned in section \ref{App}. We end with a discussion and conclusion of the results in section \ref{con}. 

For all calculations in this article the values adopted for the cosmological parameters are consistent with the latest measurements from the {\it Planck} satellite \cite{planck}: $\Omega_{\rm CDM} = 0.27$, $\Omega_b = 0.045$, $\Omega_{\kappa}=0$ (flat universe), $h=0.67$, $A_s = 2.1\times 10^{-9}$, $n_s = 0.96$. Hereby, $\Omega$ is the dimensionless energy density parameter. CDM stands for cold dark matter, $b$ stands for baryons and the matter energy density parameter is defined as $\Omega_m = \Omega_{\rm CDM} + \Omega_b$. $A_s$, $n_s$ are the amplitude and the power-law index of the primordial power spectrum, respectively. The distance unit we adopt is Mpc and the speed of light, $c$, will be set to 1.\label{cospar}

\section{Galaxy shape statistics}\label{IA&stat}

Intrinsic alignments consist of many different correlations. We focus on the correlations between the shapes of galaxies with the underlying over-density $\delta$ and velocity field $\mathbf{v}$. The over-density is defined as a dimensionless measure of the density relative to the mean density $\bar{\rho}$, $\delta(\mathbf{x}) = (\rho(\mathbf{x}) - \bar{\rho})/\bar{\rho}$. Shapes of galaxies can be described in terms of the ellipticity $e$:
\begin{equation}
    e = (1-q^2)/(1+q^2)
\end{equation}
with $q$ the ratio between the length of the minor and major axis of the best-fit ellipse of the galaxy image. The ellipticity can be decomposed in a $+$ and $\times$ component:
\begin{equation}\label{epsilon}
    e_{(+,\times)} = e[\cos(2\theta), \sin(2\theta)]
\end{equation}
with $\theta$ the angle between the x-axis and the position of the galaxy. The $+$ component describes the alignment in the radial direction relative to another galaxy when this term is negative, or in the tangential direction when this term is positive. The $\times$ term describes alignment configurations rotated by 45 degrees relative to the $+$ component. $\gamma$ is called the shear/shape and describes the change in the ellipticity induced by lensing \cite{Bernstein}, though it is also commonly adopted in alignment studies \cite[e.g.][]{Singh}. This $\gamma$ is defined as:
\begin{equation}
    \gamma = \frac{e}{2\mathcal{R}}
\end{equation}
$\mathcal{R}$ describes the sensitivity of the ellipticity to a lensing distortion.

The observed shapes of galaxies consist of multiple contributions:
\begin{equation}
    \gamma^{\rm obs} = \gamma^{\rm rnd} + \gamma^G + \gamma^I
\end{equation}
The first contribution comes from the random shapes of galaxies which is independent of the other two components. The second term is the contribution of weak lensing effects on the shapes. The last term is the effect of intrinsic alignment with the tidal field on the shapes, $\gamma^I$. This is the contribution of the shape we are interested in and from here onward we omit the superscript $I$. 

\subsection{Linear Alignment model}\label{LAmodel}

In \cite{Catelan} one of the first descriptions of the linear (tidal) alignment model was given, which describes the correlation between the intrinsic shape of galaxies and the tidal field. This model is based on assuming that when a galaxy forms, an over-dense region collapses due to its gravity, and when this happens in a constant tidal field, deformations arise. The galaxy can be stretched and therefore be aligned along the stretching axis of the tidal field or be elongated along the compressing axis for the tidal field. During galaxy evolution there are still surviving signatures of this alignment in the shape of the galaxy. 

Following \cite{Catelan}, the intrinsic shape of a galaxy can be described by two components, 

\begin{equation}\label{gamma1}
    \gamma_{(+, \times)}(\mathrm{x})=-\frac{C_{1}}{4 \pi G}\left(\nabla_{x}^{2}-\nabla_{y}^{2}, 2 \nabla_{x} \nabla_{y}\right) S\left[\Psi_{P}(\mathrm{x})\right],
\end{equation}
where $\gamma$ is related to the gravitational potential of the primordial universe $\Psi_P$, and the second derivatives describe the change in the gravitational force (hence the effect of the tidal field). $C_1$ is a normalization parameter that describes the alignment strength. Positive $C_1$ describes alignment along the stretching axis of the tidal field and negative $C_1$, alignment along its compressing axis. Galaxies tend to align with the stretching axis of the tidal field, $C_1$ is typically observed to be positive \cite{Singh, Johnston}. $S$ is a smoothing filter that smooths out fluctuations in the potential. However, since we focus on large scales ($> 10\,\mathrm{Mpc}$) this filter can be neglected \cite{Chisari}. $G$ is the gravitational constant.

On large scales, in the linear regime, the primordial gravitational potential can be connected to the over-density field, $\delta$, by means of the Poisson equation, which in Fourier space yields
\begin{equation}\label{poisson}
 \Psi_{P}(\mathbf{k})=-4 \pi G \frac{\bar{\rho}(z) a^{3}}{D(z)} k^{-2} \delta(\mathbf{k}),   
\end{equation}
and in which $\bar{\rho}(z)$ is the average density in the universe in terms of the redshift $z$, $\delta(\mathbf{k})$ is the matter over-density, $D(z)$ is the growth function that describes the growth of matter perturbations at late times, and $a$ is the scale factor the describes the expansion rate of space, with $a=1/(1+z)$. Eq. \eqref{poisson} can be substituted in the Fourier transform of Eq. \eqref{gamma1}, thus giving the relation 
\begin{equation}\label{shape}
    \gamma_{(+, \times)}(\mathbf{k})=-\:\frac{C_1 \bar{\rho}(z)}{D(z)} a^{3} \frac{(k_{x}^{2}-k_{y}^{2}, 2 k_{x} k_{y})}{k^2} \delta(\mathbf{k})
\end{equation}
Here, the model for the intrinsic shape is split in a $+$ and $\times$ contribution as in Eq. \eqref{epsilon} and Eq. \eqref{gamma1}.

When calculating the statistics with the $\times$ contribution, this turns out to be zero \cite[e.g.][]{Blazek}. This can be explained by the fact that otherwise, the $\times$ contribution would indicate a preferential rotational direction on these scales, which contradicts the cosmological principle. 

Instead of using the $+$ and $\times$ contributions for the intrinsic shape, it is more convenient to use another decomposition into $E$- and $B$-modes. This is usual in weak lensing statistics, where $E$-modes describe curl-free modes and $B$-modes describe divergence-free modes. These $E$- and $B$-modes can be calculated from the $+$ and $\times$ contributions as follows:
\begin{equation}\begin{array}{l}
\gamma_{E}(\boldsymbol{k})=\gamma_{+}(\boldsymbol{k}) \cos 2 \phi+\gamma_{\times}(\boldsymbol{k}) \sin 2 \phi \\
\gamma_{B}(\boldsymbol{k})=-\gamma_{+}(\boldsymbol{k}) \sin 2 \phi+\gamma_{\times}(\boldsymbol{k}) \cos 2 \phi
\end{array}\end{equation}
with the angle defined as $\phi=\tan ^{-1}\left(\bf{\frac{k_{x}}{k_{y}}}\right)$. The LA model only produces $E $-modes \cite{Blazek}. 

We will not go beyond the linear alignment model in this work due to limitations in the velocity reconstruction techniques. However, we note that alternative descriptions of intrinsic alignments have been provided that extend to the nonlinear regime. A widely used one is the ``nonlinear'' alignment model by \cite{Bridle}, which is a phenomenological extension of LA that replaces the linear density field in Eq. (\ref{shape}) by the nonlinear one. Alternatives are perturbation theory approaches such as those presented in \cite{Vlah, Blazekperturb}. In the long run, these could be more suitable to extend the predictions of this work consistently into the quasi-linear regime for both the alignment and velocity fields.

\subsection{Alignment statistics}

A correlation function gives the statistical correlation between two variables $X$ and $Y$ as a function of the spatial distance between them. The power spectrum is the Fourier transform of the correlation function and it can be expressed as
\begin{equation}\label{defpower}
P_{X Y}=\frac{1}{(2 \pi)^{3}} \int d^{3} \mathrm{k}\left\langle X\left(\mathrm{k}^{\prime}\right) Y(\mathrm{k})\right\rangle,
\end{equation}
with the angular brackets denoting averaging over all realizations of the variables. The auto-correlation of the matter over-density gives the matter power spectrum, 
\begin{equation}\label{Plin}
\left\langle\delta(\mathbf{k}) \delta\left(\mathbf{k}^{\prime}\right)\right\rangle=(2 \pi)^{3} P_{\delta\delta}(k) \delta^{3}\left(\mathbf{k}-\mathbf{k}^{\prime}\right)
\end{equation}
Because we restrict to linear scales, we will also call this the linear matter power spectrum.

The power spectrum of the galaxy shapes and the velocity field can be calculated by correlating the $E$-mode with the velocity field on large scales. Such velocity field can be described by a linear relation with the matter over-density,
\begin{equation}\label{velocity}
v(\mathbf{k})= \frac{f a H }{k} \delta(\mathbf{k})
\end{equation}
with $f$ the linear growth rate, and $H$, the Hubble parameter. This relation can be derived from the continuity equation in Fourier space \cite[p.270]{Dodelson}. However, when measuring the velocity field, only the component along the line of sight can be measured, also called the radial velocity:
\begin{equation}\label{vr}
v_{r}(\mathbf{k})=\frac{i k_{r}}{k} v(\mathbf{k})= \mu \frac{f a H }{k} \delta(\mathbf{k}) 
\end{equation}
where $\mu=\frac{k_{r}}{k}$ the cosine of the angle between the wave vector $\mathbf{k}$ and the line of sight. When substituting Eq. \eqref{vr} into Eq. \eqref{defpower} one can derive the cross-power spectrum between velocities and $E$-mode shapes,

\begin{equation}\label{PvrE}
P_{v_{r} E}(k, \mu, z)=-\widetilde{C}_{1}(z) f a H\left(1-\mu^{2}\right) \mu \frac{1}{k} P_{\delta \delta}(k, z).
\end{equation}
Parameter $\widetilde{C}_{1}=\frac{C_{1} \rho_{\text {crit }} \Omega_{m}}{D(z)}$ characterizes the alignment strength. A slight change of parameters is done in the terms in front of  Eq. \eqref{shape}:
\begin{equation}\Omega_{m} \rho_{\text {crit}}=\frac{\rho_{m, 0}}{\rho_{\text {crit}}} \rho_{\text {crit}}=\rho_{m} a^{3}\end{equation}
with $\rho_{\rm crit}$ the critical density, and $\rho_{m, 0}, \rho_{m}$ the matter density at present time and redshift $z$, respectively. 

The power spectrum for the galaxy density and the $E$-mode can be derived in the same way by combining Eq. \eqref{defpower} with Eq. \eqref{shape} with the expression of the galaxy over-density $\delta_g = b_g \delta$ where $b_g$ is the galaxy bias.
It is then given by
\begin{equation}\label{PgE}
P_{g E}(k, \mu, z)=-\widetilde{C}_{1} b_{g} \left(1-\mu^{2}\right) \left(1+\frac{f}{b_g} \mu^2\right) P_{\delta \delta}(k).
\end{equation}
Here, the extra factor of $(1+\frac{f}{b_g} \mu^2)$ comes from the change in the galaxy over-density due to redshift space distortions (RSD), which influence the measured clustering of galaxies on redshift space because of contribution from the movement of the galaxies to the observed redshift (see e.g. Appendix A of \cite{Singh}).

The angular information of these power spectra can be best extracted through the multipole expansion of Eq. \eqref{PvrE} and Eq. \eqref{PgE} \cite{Okumura2}. For a generic power spectrum, this is given by
\begin{equation}\label{multipole}
P_{XY}^{(\ell)}(k) \equiv \frac{2 \ell+1}{2} \int_{-1}^{1} \mathrm{d} \mu \mathcal{L}_{\ell}(\mu) P_{XY}(k, \mu),
\end{equation}
with $\ell$ the order of the moment and $\mathcal{L}$ the Legendre polynomial of order $\ell$.

\section{Reconstruction of the velocity field}\label{vrec}

A method to calculate the velocity field from the data is required for executing the prediction of the detectability of the velocity-shape correlation. There are multiple ways to measure the velocity field, for example, with peculiar velocities \cite{Strauss}. Here, we focus on a reconstruction of the large-scale velocity field as provided by the kinetic Sunyaev-Zeldovich effect (kSZ effect) \cite{Sunyaev, Hand}.

The kSZ effect is based on the following process. 
Similarly to galaxies, free electrons in the Universe also inherit a peculiar velocity complement to their thermal movements. Photons from the Cosmic Microwave Background (CMB) are scattered off these electrons; this process is called inverse Compton scattering. Because of the peculiar velocity of the electrons, the scattering is not elastic and an energy shift takes place. The photons are therefore Doppler shifted, which can be measured by looking at the CMB temperature perturbation spectrum. When comparing the CMB spectrum with an overlapping galaxy survey, the peculiar velocity from the electrons and in general, matter, can be reconstructed. 

Ref. \cite{Smith} shows that measuring the kSZ bispectrum can provide an estimate for the radial velocity field and the velocity noise spectrum (see section III E of that work), both expressed in terms of the electron density-galaxy density power spectrum, galaxy density-galaxy density power spectrum and the CMB spectrum. Furthermore, the radial velocity depends exponentially on the optical depth parameter $\tau$ for which the fiducial value of $\tau=0.06$ is adopted, consistently with the measurements of the {\it Planck} satellite \cite{planck}. However, in the linear limit, the reconstruction of the radial velocity converges to the linear relation of Eq. \eqref{vr} up to a factor $b_v$ which is not known in advance and must be marginalized over:
\begin{equation}
\label{bias}
\hat{v}_{r}=b_{v} v_{r}.
\end{equation}
This parameter appears because of a degeneracy with the optical depth \cite{Smith}. Such degeneracy can be broken by using the cross-correlation between the reconstructed radial velocity field and the galaxy density $P_{\hat{v}_rg}$ \cite{Sugiyama, Smith}, for example, or by relying on additional CMB observables. In the case of the galaxy alignment bias, the degeneracy can be broken by adding the position-intrinsic shape power spectrum to the data vector.

Notice that this specific method for the reconstruction of the radial velocity field is intrinsically limited to linear scales (Eq. \ref{vr}), where it has been suggested to perform better than galaxy surveys per se \cite{Smith}. For this reason, our predictions are restricted to $k_{\rm max}<1$ Mpc$^{-1}$. This also justifies adopting the LA model as our baseline in Section \ref{LAmodel}. In Section \ref{FF}, we explore the impact of restricting $k_{\rm max}$ to even smaller values.

For predicting the noise spectrum $N_{\hat{v}_r}$ of the reconstructed radial velocity field, the approximation proposed by \cite{Zhu} can be used:
\begin{equation}\label{noise}
N_{\hat{v}_{r}} /(a f H)^{2}=10^{6} h^{-5} \mathrm{Mpc}^{5},
\end{equation}
with the approximate condition that the number density of the used galaxy survey for the prediction of the detectability will be $\bar{n}_{\rm gal}=10^{-3}\,h^{3}\, \mathrm{Mpc}^{-3}$ or $\bar{n}_{\rm gal}=10^{-4}\,h^{3}\, \mathrm{Mpc}^{-3}$ \label{condition}. This is in line with our choice of spectroscopic sample, further described in Section \ref{FF}. Furthermore, when adopting this noise spectrum, one is implicitly assuming that the kSZ measurement will have such quality measurements as expected from the Simons Observatory \cite{Simonobs}. This corresponds to a noise level of 6 $\mu$K-arcmin for a survey with 1.5 arcmin beam. Finally, we assume that the kSZ bispectrum measurement is free from contamination from thermal SZ or the cosmic infrared background, as the former is parity-odd and the latter are parity-even. 

\section{Fisher forecast and results}\label{FF}

Our prediction for the detectability of the signal is done by means of a Fisher forecast \cite{Tegmark,Coe}. 
When a Fisher matrix is calculated for multiple model parameters, the inverse of this matrix gives the covariance matrix for these parameters. The Fisher matrix for cosmological parameters is given by \cite{OkumuraTaruya}
\begin{equation}\label{Fisher}
F_{i, j}=\frac{V_{\text {survey}}}{(2 \pi)^{2}} \int_{k_{\min }}^{k_{\max }} d k\,k^{2} \int_{-1}^{1} d \mu \sum_{X, Y} \frac{\partial P_{X}(k, \mu)}{\partial \theta_{i}} \frac{1}{\sigma_{XY}} \frac{\partial P_{Y}(k, \mu)}{\partial \theta_{j}}.
\end{equation}
$P_X$, $P_Y$ are the power spectra of interest that contain the cosmological information of the model. $\theta_i$ are the parameters of which the constrains are required. 

In this case, the detectability of the velocity-shape correlation needs to be measured so both $X$, $Y$ will be $\hat{v}_rE$. For comparison the same calculation for $X$, $Y$ being $gE$ is done, this is the correlation between the galaxy density and the shapes and has already been measured in the data \cite{Singh}. Here, we will look at parameter $\widetilde{C}_{1}$. Tight constrains on this parameter indicate a good detectability of the alignment, hence also of the velocity-shape correlation. $\sigma_{XY}$ is the covariance matrix of the power spectra. In our case, when focussing on only one parameter,
this collapses to a scalar. Because this calculation is done for large scales, a Gaussian covariance can be assumed \cite{Taruya}. 

We perform the Fisher calculation of Eq. \eqref{Fisher} for the multipole power spectrum and the corresponding covariance terms using the expansion in Eq. \eqref{multipole} as follows. 

\begin{itemize}

\item We consider the dipole term of $P_{\hat{v}_rE}$ in Eq. \eqref{PvrE} with velocity bias from Eq. \eqref{bias}, and the monopole term of $P_{gE}$ in Eq. \eqref{PgE}. These contain most of the information \cite{Okumura2} and will therefore be the power spectra of interest. They are given by
\begin{equation}\label{multipoleexp}
\begin{array}{l}
P_{\hat{v}_{r} E}^{(1)}(k) \equiv \frac{3}{2} \int_{-1}^{1} \mathrm{d} \mu \mathcal{L}_{1}(\mu) P_{\hat{v}_{r} E}(k, \mu) \\
P_{g E}^{(0)}(k) \equiv \frac{1}{2} \int_{-1}^{1} \mathrm{d} \mu \mathcal{L}_{0}(\mu) P_{g E}(k, \mu)
\end{array}
\end{equation}

\item The Gaussian covariance values for $P_{\hat{v}_rE}$ and $P_{gE}$ in the multipole formalism are described in \cite[page 11]{kurita}:
\begin{equation}\label{mpcov}\begin{array}{c}
\mathrm{C}_{11}^{P_{v_r E}^{(1)}}=\sigma_{P_{\hat{v}_rE}^{(1)}, P_{\hat{v}_rE}^{(1)}}^{2} \equiv 9 \int_{-1}^{1} d \mu \mathcal{L}_{1}(\mu) \mathcal{L}_{1}(\mu) \\
\times\left[P_{\hat{v}_rE}^{2}(k, \mu)+\widetilde{P}_{\hat{v}_r\hat{v}_r}(k, \mu) \widetilde{P}_{E E}(k, \mu)\right] \\
\mathrm{C}_{00}^{P_{g E}^{(0)}}=\sigma_{P_{g E}^{(1)}, P_{g E}^{(1)}}^{2} \equiv \int_{-1}^{1} d \mu \mathcal{L}_{0}(\mu) \mathcal{L}_{0}(\mu) \\
\times\left[P_{g E}^{2}(k, \mu)+\widetilde{P}_{g g}(k, \mu) \widetilde{P}_{E E}(k, \mu)\right]
\end{array}\end{equation}
where $\tilde{P}_{gg} = P_{gg} + \frac{1}{\bar{n}_{gal}}$, $\tilde{P}_{\hat{v}_r\hat{v}_r} = P_{\hat{v}_r\hat{v}_r} +N_{v_r}$ and $\tilde{P}_{EE} = P_{EE} + \frac{\sigma_{\gamma}}{\bar{n}_{gal}}$ are the power spectra with noise terms. $\bar{n}_{\mathrm{gal}}$ is the mean number of galaxies incorporating the shot noise, $N_{\hat{v}_r}$, the velocity noise spectrum approximated by Eq. \eqref{noise}, and $\sigma_{\gamma}$, the scatter in the intrinsic shape arising due to the shape noise.

\item The Fisher calculation of Eq. \eqref{Fisher} in multipole formalism is given by
\begin{equation}\label{mpFisher}\begin{array}{l}
F_{\bar{C}_{1} \tilde{C}_{1}}=\frac{V_{\text {survey }}}{(2 \pi)^{2}} \int_{k_{\text {min }}}^{k_{\max }} d k \,k^{2}\left(\frac{\partial P_{\hat{v}_r E}^{(1)}(k)}{\partial \widetilde{C}_1}\right)^{2} \frac{1}{\sigma^2_{P_{\hat{v}_rE}^{(1)}, P_{\hat{v}_rE}^{(1)}}} \\
F_{\tilde{C}_{1} \tilde{C}_{1}}=\frac{V_{\text {survey }}}{(2 \pi)^{2}} \int_{k_{\text {min }}}^{k_{\text {max }}} d k\,k^{2}\left(\frac{\partial P_{g E}^{(0)}(k)}{\partial \widetilde{C}_1}\right)^{2} \frac{1}{\sigma^2_{P_{g E}^{(0)}, P_{g E}^{(0)}}}.
\end{array}\end{equation}
The square root of the inverse of this expression gives the uncertainty of the alignment strength parameter $\widetilde{C}_1$.\label{uncertainty}

\end{itemize}

We execute our predictions for a combination of the upcoming LSST \cite{LSST} and 4MOST \cite{4most} galaxy surveys. We focus on the Luminous Red Galaxy sample expected to be delivered by the 4MOST Cosmology Redshift Survey \cite{Richard}.
The combination is advantageous in that LSST would provide accurate galaxy shapes, and 4MOST, complementary spectroscopic redshift measurements.
This is a reasonable assumption, given that 4MOST targets are expected to be selected from much shallower precursor surveys.
We assume full overlap between LSST, 4MOST and the Simons Observatory southern survey footprints. In practice, the final area coverage ($7,500$ sq. deg.) is restricted by the 4MOST data set.

The accuracy of the spectroscopic redshift measurements is needed to be able to separate the shape altering effects of alignments from the weak lensing effects. With this information, one can select the galaxies around the same redshift, ruling out the possibility of a lensing effect between those galaxies. 
Our assumptions about the data used for the Fisher forecast are shown in Table \ref{table1}. All predictions make use of the publicly available Core Cosmology Library \cite{ccl}, version 2.1. 

\begin{table}[h!]
\begin{tabular}{lllllll}
\cline{1-5}
\multicolumn{1}{|l|}{} & \multicolumn{1}{l|}{$\langle z \rangle$} & \multicolumn{1}{l|}{\textbf{$\bar{n}_{\rm gal} (\mathrm{Mpc}^{-3})$}} & \multicolumn{1}{l|}{\textbf{$V_{\rm survey}(\mathrm{Mpc}^3)$}} & \multicolumn{1}{l|}{$\widetilde{C}_{1}$} & \textbf{} &  \\ \cline{1-5}
\multicolumn{1}{|l|}{4MOST + LSST} & \multicolumn{1}{l|}{0.55} & \multicolumn{1}{l|}{\textbf{$2.9 \times 10^{-4}$}} & \multicolumn{1}{l|}{$0.10\times 10^{11}$} & \multicolumn{1}{l|}{0.026} & \textbf{} &  \\ \cline{1-5}
\end{tabular}
\caption{The considered properties of the combination of the 4MOST and LSST galaxy surveys for the Fisher forecast.}\label{table1}
\end{table}
The average redshift denotes the mean depth that the survey probes. The predicted data from the 4MOST survey \cite{4most} is compatible with the requirements (\ref{condition}) for using Eq. \eqref{noise} as $N_{\hat{v}_r}$. We integrate the S/N between $k_{\min }=2 \pi / V_{\text {\rm survey}}^{\frac{1}{3}}\: \mathrm{Mpc}^{-1}$ and $k_{\max }=\{0.2,\: 0.5,\: 1\:\}\mathrm{Mpc}^{-1}$. Adopting different $k_{\rm max}$ values allows us to quantify how much information is available in the mildly nonlinear regime. Following \cite{Taruya}, we take $\sigma_{\gamma} = 0.3$ for the typical shape noise. $b_v$ of the velocity reconstruction is set to $1$, but in general it should be predicted by means of numerical simulations or marginalized over. 

\begin{figure}[t!]
\includegraphics[width = 0.5\textwidth]{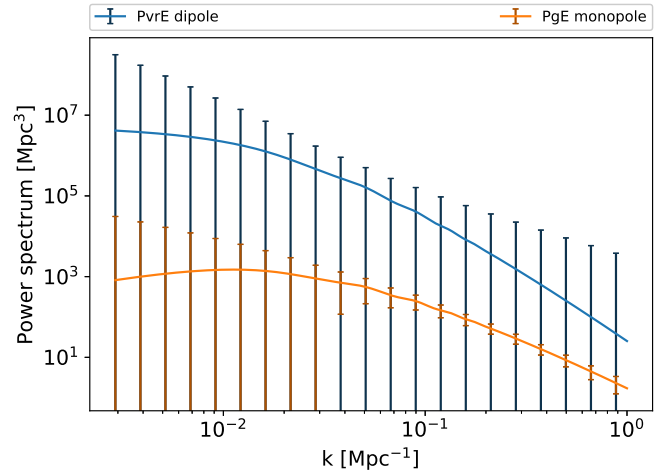}
\caption{Dipole $P_{\hat{v}_rE}$ (blue) and monopole $P_{gE}$ (orange) as a function of wave-number for a value of $k_{\max}=1$, with accompanying error bars corresponding to the 4MOST+LSST combination. The velocity reconstruction is assumed to be obtained from the Simons Observatory kSZ measurements.}
\label{plot}
\end{figure}

From the Fisher forecast executed by calculating Eq. \eqref{mpFisher}, the uncertainty of the alignment strength $\widetilde{C}_1$ parameter is given by square root of the inverse of the Fisher value $F_{\bar{C}_{1} \tilde{C}_{1}}$. 
This is divided by the value of $\widetilde{C}_1$ to get the fractional error on this parameter. 
The dipole of $P_{\hat{v}_rE}$ and the monopole of $P_{gE}$ with their uncertainties are shown in Figure \ref{plot}. The signal-to-noise ratio (S/N ratio) of these power spectra can be calculated \cite[Eq.   25]{kurita}. The results of the forecast are shown in Table \ref{table2}.

\begin{table}[]
\begin{tabular}{|l|c|c|c|c}
\cline{1-4}
Correlation & $k_{\rm max}$ [Mpc$^{-1}$] & $\delta\widetilde{C}_{1}/\widetilde{C}_{1}$ & S/N ratio &  \\ \cline{1-4}\cline{1-4}
\multirow{3}{*}{$P_{\hat{v}_{r}E}^{(1)}$} & 0.2 & 0.19 & 7.4 &  \\ \cline{2-4}
 & 0.5 & 0.19 & 7.4 &  \\ \cline{2-4}
 & 1 & 0.19 & 7.4 &  \\ \cline{1-4}
\multirow{3}{*}{$P_{gE}^{(0)}$} & 0.2 & 0.032 & 44 &  \\ \cline{2-4}
 & 0.5 & 0.023 & 62 &  \\ \cline{2-4}
 & 1 & 0.021 & 69 &  \\ \cline{1-4}
\end{tabular}
\caption{Results of the Fisher forecast with 4MOST+LSST on $P_{\hat{v}_rE}^{(1)}$ and $P_{gE}^{(0)}$ to parameter $\widetilde{C}_1$. The velocity reconstruction is assumed to be obtained from the Simons Observatory kSZ bispectrum measurements as proposed in \cite{Smith}.}\label{table2}
\end{table}

We also carried out S/N predictions for higher order poles of $P_{\hat{v}_rE}$ and $P_{gE}$ by computing the multipole expansions of Eq. \eqref{multipoleexp} and the covariances of Eq. \eqref{mpcov} for higher poles of the Legendre polynomials as in Eq. \eqref{multipole}. By symmetry, only the odd poles are nonzero for $P_{\hat{v}_rE}$, and only the even ones for $P_{gE}$. The S/N ratios for the third order pole of $P_{\hat{v}_rE}$ (5.0, regardless of $k_{\max}$) and the second order pole of $P_{gE}$ (respectively: 16, 23, 25 for increasing $k_{\max}$) are still significant. For higher orders, the S/N becomes negligible. However, the dipole and monopole have the highest S/N ratios for every value of $k_{\max}$.

As can be seen from Table \ref{table2}, the fractional errors of the velocity-shape correlation are smaller than $1$ for the different values of $k_{\max}$. They are also larger than the fractional errors for the corresponding $k_{\max}$ for the galaxy density-shape correlation. Overall, Table \ref{table2} and Figure \ref{plot} allow us to conclude that the S/N ratios are significant and consistent with the fractional error estimates.

The S/N ratio is still significant for the lower value of $k_{\max}=0.2$ Mpc$^{-1}$. We can thus predict that a detection of the velocity-shape correlation signal in the data from these upcoming galaxy surveys is indeed possible.

\section{Applications}\label{App}
The significant results of section \ref{FF} indicate that measuring the velocity-shape correlation signal is well-within the possibilities of soon-to-come experiments. In this section, we discuss two different potential applications of such measurement.

\subsection{Mitigating selection effects induced by intrinsic alignments on spectroscopic surveys}

\begin{figure}[h!]
\centering
\includegraphics[scale = 0.4]{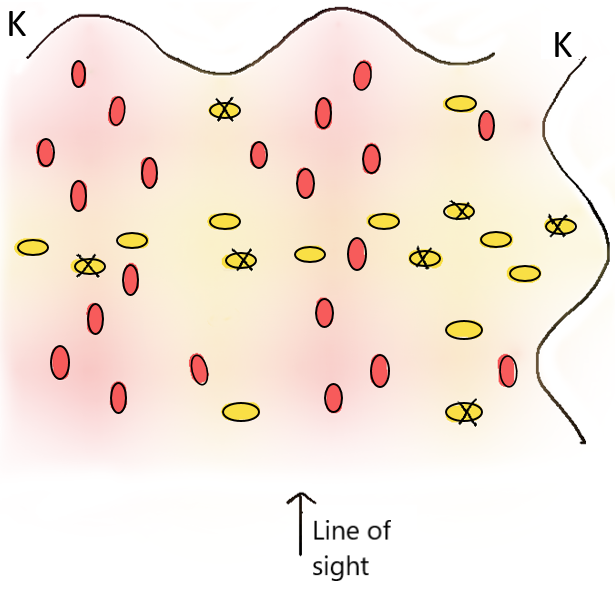}
\caption{Schematic picture of the observational selection effect. The lower probability of measuring perpendicular galaxy is depicted with the exaggerated crossing out of some of these galaxies. Figure based on \cite{Martens}. }
\label{fig:cartoon}
\end{figure}

In \cite{Hirata,Krause11,Martens}, it is shown that the intrinsic alignments of galaxies induce a selection effect in spectroscopic surveys. Galaxies aligned along the line of sight are more likely to be observed than galaxies aligned perpendicularly to the line of sight. When a galaxy is aligned along the line of sight, its stars are packed together in a smaller observed area, making it more likely to be observed. The alignments of galaxies correlate with the large scale structure shown in Eq. \eqref{PgE}
and hence so does this selection effect. The smaller probability of measuring the perpendicularly aligned galaxies is sketched in Figure \ref{fig:cartoon} as a crossing-out of galaxies. The measured Fourier modes drawn above and right in the picture are altered due to this selection. This alteration mimics the change in Fourier modes due to clustering of galaxies because of RSD effects, making it harder to tell the selection and RSD effects apart. The ability to measure the selection effect makes it possible to correct for the alterations in the Fourier modes and mitigate the effects of selection when measuring RSDs. 

In \cite{Martens}, they measure the selection effect by splitting their galaxy sample in galaxies oriented parallel and perpendicular to the line of sight, and they find a significant difference in the measurement of clustering, which is attributed to their intrinsic alignment. Their findings are in good agreement with the theoretical description of the effect by \cite{Hirata}. 
In such description, an extra term, $\epsilon$, appears in the galaxy over-density field. The galaxy over-density at linear order is thus
\begin{equation}\delta_{g}({\bf k}) = (b + f\mu^2)\delta_{m}({\bf k}) + \epsilon({\bf \hat{e}_3} \mid {\bf k}), 
\end{equation} 
with ${\bf \hat{e}_3}$ the line of sight direction.

In the linear limit, the intrinsic alignments follow the tidal field described in Eq. \eqref{gamma1} 
and $\epsilon$ becomes
\begin{equation}
\epsilon({\bf \hat{n}} \mid {\bf x}) = A s_{ij}({\bf x})\hat{n}_i\hat{n}_j,
\end{equation}
with $A$ a bias parameter that depends on intrinsic alignment and the selection effects, $\hat{n}_{i}$ the component of directional unit vector and $s_{ij}$ the dimensionless tidal field; $s_{i j}(\mathbf{x})=\left[\nabla_{i} \nabla_{j} \nabla^{-2}-\frac{1}{3} \delta_{i j}\right] \delta_{\mathrm{m}}(\mathbf{x})$. 

The galaxy density power spectrum changes from:
\begin{equation}
P_{gg}(k, \mu, z)=\left(b_g+f\mu^{2}\right)^{2} P_{\delta\delta}(k)
\end{equation}
to:
\begin{equation}
P_{gg}(k, \mu, z)=\left[b_g-\frac{A}{3}+(f+A) \mu^{2}\right]^{2} P_{\delta\delta}(k).
\end{equation}
Alignments can be interpreted as a change in the RSD term with $f\rightarrow f+A$. 

Notice that this result implies that both the galaxy-shape and velocity-shape power spectra contain these extra selection terms, appearing as
\begin{equation}
P_{gE}(k, \mu, z)=-\widetilde{C}_1 (1-\mu^2)\left[b_g-\frac{A}{3}+(f+A) \mu^{2}\right] P_{\delta\delta}(k) 
\end{equation}
and
\begin{equation}
P_{g\hat{v}_r}(k, \mu, z) =\left[b_g-\frac{A}{3}+(f+A) \mu^{2}\right] b_v f a Hk^{-1} \mu P_{\delta\delta}(k).
\end{equation}
In contrast, neither the velocity shape correlation of Eq. \eqref{PvrE}, nor the shape-shape correlation:
\begin{equation}P_{E E}(k, \mu, z)=\left[\widetilde{C}_{1}(z)\left(1-\mu^{2}\right)\right]^{2} P_{\delta \delta}(k, z)
\end{equation}
nor velocity-velocity correlation:
\begin{equation}P_{\hat{v}_r\hat{v}_r}(k, \mu, z)=\left(\frac{\mu b_{v} f a H}{k}\right)^{2} P_{\delta \delta}(k, z)\end{equation}
which only depend on the dark matter over-density at linear order (and not on the the biased galaxy density), contain the selection effect parameter $A$. 

Overall, the power spectra $P_{gg}$, $P_{gE}$ and $P_{g\hat{v}_r}$ contain this extra parameter $A$ but in linear approximation $P_{\hat{v}_rE}$, $P_{EE}$ and $P_{\hat{v}_r\hat{v}_r}$ do not. Combining measurements from these different correlations can in principle break the degeneracies in determining this selection parameter $A$.  

\subsection{Scale dependence of intrinsic alignment}
\label{sec:ii}

Intrinsic alignment effects can be well described on large scales by the linear alignment model (\ref{LAmodel}). 
However, on small scales, accurate models are still under development \cite{Singh, Fortuna}. The scale dependence of intrinsic alignment at these smallest scales can be better determined by using multiple measurements of the galaxy shapes \cite{Leonard}. It was already found in \cite{Singh&Mandelbaum} that different shape measurements of a galaxy lead to a scale-independent multiplicative change in the alignment amplitude. The outer part of a galaxy is more sensitive to alignment effects due to the weaker gravitational force than the inner parts of the galaxy \cite{Georgiou, Tenneti, Velliscig, Chisarisim}. 

Ref. \cite{Leonard} proposed the use of two different shape measurements for constraining the scale-dependence of intrinsic alignments at small scales. They found that this method improved on already existing methods when the difference between the two shape measurements is large and when the correlation between the shape noise and the shape measurements is also large.

We perform a Fisher forecast with the same galaxy sample and kSZ measurement as in section \ref{FF}, to determine whether the addition of the velocity-shape correlation could aid further in determining scale-dependence of intrinsic alignments. 
We introduce a parameter $\alpha$ that describes the change in the alignment signal to linear order:
\begin{equation}
\begin{array}{l}
\Delta P_{\hat{v}_{r} E}=P_{\hat{v}_{r} E, 1}-P_{\hat{v}_{r} E, 2}=(1-\alpha) P_{\hat{v}_{r} E} \\
\Delta P_{g E}=P_{g E, 1}-P_{g E, 2}=(1-\alpha) P_{gE}
\end{array}
\end{equation}
with the subscripts $1,2$ denoting the power spectra for different shape measurements. This prediction is again done in the linear regime, therefore we choose $k_{\max}=0.2$ Mpc$^{-1}$ to mitigate the nonlinear effects.

We find an improvement of the fractional error of $\alpha$ from 0.040 without velocity-shape correlation to 0.026 with velocity-shape correlation. Notice that the parameter is overall multiplicative and does not introduce additional scale-dependence.
As a first estimate, adding velocity-shape correlation to constrain the multiple shape measurements is promising. This is in line with our results from Table \ref{table2}. The next step would be to revisit these conclusions in the context of modelling extensions that include either a phenomenological or physically motivated (e.g. perturbative) scale dependence.

\section{Conclusion and discussion}\label{con}

We performed a Fisher forecast for the detectability of the velocity-intrinsic shape correlation for the combination of 4MOST+LSST, where the velocity reconstruction is assumed to be obtained from the kSZ effect bispectrum measured by the Simons Observatory. We find that such correlation could be observed in the next decade. 

The combination of 4MOST+LSST allows us to separate the weak lensing effects from the alignment effects. The calculation could be extended in the future to predict the joint measurement of alignments and lensing from LSST alone. This would require incorporation of photometric reshift uncertainties, but it would be advantageous in terms of a larger area. 

Because of the possibility of detection, we looked at two possible applications for the velocity-shape correlation. Implementing velocity-shape correlation to break degeneracies between RSD and alignment effects on galaxy clustering measurements is promising. It would be interesting to perform a prediction on the potential gains. Measuring the velocity-shape correlation with multiple shape measurements of galaxies would also allow for gains in determining the scale-dependence of intrinsic alignments. This encourages further estimations on the possible improvements at higher order (i.e., beyond the linear alignment model and linear velocity field reconstruction). 

\section{Acknowledgements}

We thank Teppei Okumura, Karel Zwetsloot and the anonymous referee for feedback that helped improved this manuscript. This work is part of the Delta ITP consortium, a program of the Netherlands Organisation for Scientific Research (NWO) that is funded by the Dutch Ministry of Education, Culture and Science (OCW).

\bibliography{references}

\providecommand{\noopsort}[1]{}\providecommand{\singleletter}[1]{#1}%
\begin{thebibliography}{51}%
\makeatletter
\providecommand \@ifxundefined [1]{%
 \@ifx{#1\undefined}
}%
\providecommand \@ifnum [1]{%
 \ifnum #1\expandafter \@firstoftwo
 \else \expandafter \@secondoftwo
 \fi
}%
\providecommand \@ifx [1]{%
 \ifx #1\expandafter \@firstoftwo
 \else \expandafter \@secondoftwo
 \fi
}%
\providecommand \natexlab [1]{#1}%
\providecommand \enquote  [1]{``#1''}%
\providecommand \bibnamefont  [1]{#1}%
\providecommand \bibfnamefont [1]{#1}%
\providecommand \citenamefont [1]{#1}%
\providecommand \href@noop [0]{\@secondoftwo}%
\providecommand \href [0]{\begingroup \@sanitize@url \@href}%
\providecommand \@href[1]{\@@startlink{#1}\@@href}%
\providecommand \@@href[1]{\endgroup#1\@@endlink}%
\providecommand \@sanitize@url [0]{\catcode `\\12\catcode `\$12\catcode
  `\&12\catcode `\#12\catcode `\^12\catcode `\_12\catcode `\%12\relax}%
\providecommand \@@startlink[1]{}%
\providecommand \@@endlink[0]{}%
\providecommand \url  [0]{\begingroup\@sanitize@url \@url }%
\providecommand \@url [1]{\endgroup\@href {#1}{\urlprefix }}%
\providecommand \urlprefix  [0]{URL }%
\providecommand \Eprint [0]{\href }%
\providecommand \doibase [0]{http://dx.doi.org/}%
\providecommand \selectlanguage [0]{\@gobble}%
\providecommand \bibinfo  [0]{\@secondoftwo}%
\providecommand \bibfield  [0]{\@secondoftwo}%
\providecommand \translation [1]{[#1]}%
\providecommand \BibitemOpen [0]{}%
\providecommand \bibitemStop [0]{}%
\providecommand \bibitemNoStop [0]{.\EOS\space}%
\providecommand \EOS [0]{\spacefactor3000\relax}%
\providecommand \BibitemShut  [1]{\csname bibitem#1\endcsname}%
\let\auto@bib@innerbib\@empty
\bibitem [{\citenamefont {{Catelan}}\ \emph {et~al.}(2001)\citenamefont
  {{Catelan}}, \citenamefont {{Kamionkowski}},\ and\ \citenamefont
  {{Blandford}}}]{Catelan}%
  \BibitemOpen
  \bibfield  {author} {\bibinfo {author} {\bibfnamefont {P.}~\bibnamefont
  {{Catelan}}}, \bibinfo {author} {\bibfnamefont {M.}~\bibnamefont
  {{Kamionkowski}}}, \ and\ \bibinfo {author} {\bibfnamefont {R.~D.}\
  \bibnamefont {{Blandford}}},\ }\href {\doibase
  10.1046/j.1365-8711.2001.04105.x} {\bibfield  {journal} {\bibinfo  {journal}
  {\mnras}\ }\textbf {\bibinfo {volume} {320}},\ \bibinfo {pages} {L7}
  (\bibinfo {year} {2001})},\ \Eprint {http://arxiv.org/abs/astro-ph/0005470}
  {arXiv:astro-ph/0005470 [astro-ph]} \BibitemShut {NoStop}%
\bibitem [{\citenamefont {{Brown}}\ \emph {et~al.}(2002)\citenamefont
  {{Brown}}, \citenamefont {{Taylor}}, \citenamefont {{Hambly}},\ and\
  \citenamefont {{Dye}}}]{Brown}%
  \BibitemOpen
  \bibfield  {author} {\bibinfo {author} {\bibfnamefont {M.~L.}\ \bibnamefont
  {{Brown}}}, \bibinfo {author} {\bibfnamefont {A.~N.}\ \bibnamefont
  {{Taylor}}}, \bibinfo {author} {\bibfnamefont {N.~C.}\ \bibnamefont
  {{Hambly}}}, \ and\ \bibinfo {author} {\bibfnamefont {S.}~\bibnamefont
  {{Dye}}},\ }\href {\doibase 10.1046/j.1365-8711.2002.05354.x} {\bibfield
  {journal} {\bibinfo  {journal} {\mnras}\ }\textbf {\bibinfo {volume} {333}},\
  \bibinfo {pages} {501} (\bibinfo {year} {2002})},\ \Eprint
  {http://arxiv.org/abs/astro-ph/0009499} {arXiv:astro-ph/0009499 [astro-ph]}
  \BibitemShut {NoStop}%
\bibitem [{\citenamefont {{Mandelbaum}}\ \emph {et~al.}(2006)\citenamefont
  {{Mandelbaum}}, \citenamefont {{Hirata}}, \citenamefont {{Ishak}},
  \citenamefont {{Seljak}},\ and\ \citenamefont
  {{Brinkmann}}}]{MandelbaumHirata}%
  \BibitemOpen
  \bibfield  {author} {\bibinfo {author} {\bibfnamefont {R.}~\bibnamefont
  {{Mandelbaum}}}, \bibinfo {author} {\bibfnamefont {C.~M.}\ \bibnamefont
  {{Hirata}}}, \bibinfo {author} {\bibfnamefont {M.}~\bibnamefont {{Ishak}}},
  \bibinfo {author} {\bibfnamefont {U.}~\bibnamefont {{Seljak}}}, \ and\
  \bibinfo {author} {\bibfnamefont {J.}~\bibnamefont {{Brinkmann}}},\ }\href
  {\doibase 10.1111/j.1365-2966.2005.09946.x} {\bibfield  {journal} {\bibinfo
  {journal} {\mnras}\ }\textbf {\bibinfo {volume} {367}},\ \bibinfo {pages}
  {611} (\bibinfo {year} {2006})},\ \Eprint
  {http://arxiv.org/abs/astro-ph/0509026} {arXiv:astro-ph/0509026 [astro-ph]}
  \BibitemShut {NoStop}%
\bibitem [{\citenamefont {{Joachimi}}\ \emph {et~al.}(2011)\citenamefont
  {{Joachimi}}, \citenamefont {{Mandelbaum}}, \citenamefont {{Abdalla}},\ and\
  \citenamefont {{Bridle}}}]{Joachimi}%
  \BibitemOpen
  \bibfield  {author} {\bibinfo {author} {\bibfnamefont {B.}~\bibnamefont
  {{Joachimi}}}, \bibinfo {author} {\bibfnamefont {R.}~\bibnamefont
  {{Mandelbaum}}}, \bibinfo {author} {\bibfnamefont {F.~B.}\ \bibnamefont
  {{Abdalla}}}, \ and\ \bibinfo {author} {\bibfnamefont {S.~L.}\ \bibnamefont
  {{Bridle}}},\ }\href {\doibase 10.1051/0004-6361/201015621} {\bibfield
  {journal} {\bibinfo  {journal} {\aap}\ }\textbf {\bibinfo {volume} {527}},\
  \bibinfo {eid} {A26} (\bibinfo {year} {2011})},\ \Eprint
  {http://arxiv.org/abs/1008.3491} {arXiv:1008.3491 [astro-ph.CO]} \BibitemShut
  {NoStop}%
\bibitem [{\citenamefont {Singh}\ \emph {et~al.}(2015)\citenamefont {Singh},
  \citenamefont {Mandelbaum},\ and\ \citenamefont {More}}]{Singh}%
  \BibitemOpen
  \bibfield  {author} {\bibinfo {author} {\bibfnamefont {S.}~\bibnamefont
  {Singh}}, \bibinfo {author} {\bibfnamefont {R.}~\bibnamefont {Mandelbaum}}, \
  and\ \bibinfo {author} {\bibfnamefont {S.}~\bibnamefont {More}},\ }\href
  {\doibase 10.1093/mnras/stv778} {\bibfield  {journal} {\bibinfo  {journal}
  {Monthly Notices of the Royal Astronomical Society}\ }\textbf {\bibinfo
  {volume} {450}},\ \bibinfo {pages} {2195} (\bibinfo {year}
  {2015})}\BibitemShut {NoStop}%
\bibitem [{\citenamefont {{Johnston}}\ \emph {et~al.}(2019)\citenamefont
  {{Johnston}} \emph {et~al.}}]{Johnston}%
  \BibitemOpen
  \bibfield  {author} {\bibinfo {author} {\bibfnamefont {H.}~\bibnamefont
  {{Johnston}}} \emph {et~al.},\ }\href {\doibase 10.1051/0004-6361/201834714}
  {\bibfield  {journal} {\bibinfo  {journal} {\aap}\ }\textbf {\bibinfo
  {volume} {624}},\ \bibinfo {eid} {A30} (\bibinfo {year} {2019})},\ \Eprint
  {http://arxiv.org/abs/1811.09598} {arXiv:1811.09598 [astro-ph.CO]}
  \BibitemShut {NoStop}%
\bibitem [{\citenamefont {{Samuroff}}\ \emph {et~al.}(2019)\citenamefont
  {{Samuroff}} \emph {et~al.}}]{Samuroff}%
  \BibitemOpen
  \bibfield  {author} {\bibinfo {author} {\bibfnamefont {S.}~\bibnamefont
  {{Samuroff}}} \emph {et~al.} (\bibinfo {collaboration} {DES collaboration}),\
  }\href {\doibase 10.1093/mnras/stz2197} {\bibfield  {journal} {\bibinfo
  {journal} {\mnras}\ }\textbf {\bibinfo {volume} {489}},\ \bibinfo {pages}
  {5453} (\bibinfo {year} {2019})},\ \Eprint {http://arxiv.org/abs/1811.06989}
  {arXiv:1811.06989 [astro-ph.CO]} \BibitemShut {NoStop}%
\bibitem [{\citenamefont {{Mandelbaum}}\ \emph {et~al.}(2011)\citenamefont
  {{Mandelbaum}} \emph {et~al.}}]{Mandelbaum}%
  \BibitemOpen
  \bibfield  {author} {\bibinfo {author} {\bibfnamefont {R.}~\bibnamefont
  {{Mandelbaum}}} \emph {et~al.},\ }\href {\doibase
  10.1111/j.1365-2966.2010.17485.x} {\bibfield  {journal} {\bibinfo  {journal}
  {\mnras}\ }\textbf {\bibinfo {volume} {410}},\ \bibinfo {pages} {844}
  (\bibinfo {year} {2011})},\ \Eprint {http://arxiv.org/abs/0911.5347}
  {arXiv:0911.5347 [astro-ph.CO]} \BibitemShut {NoStop}%
\bibitem [{\citenamefont {{Hirata}}\ \emph {et~al.}(2007)\citenamefont
  {{Hirata}}, \citenamefont {{Mandelbaum}}, \citenamefont {{Ishak}},
  \citenamefont {{Seljak}}, \citenamefont {{Nichol}}, \citenamefont
  {{Pimbblet}}, \citenamefont {{Ross}},\ and\ \citenamefont
  {{Wake}}}]{NLAhirata}%
  \BibitemOpen
  \bibfield  {author} {\bibinfo {author} {\bibfnamefont {C.~M.}\ \bibnamefont
  {{Hirata}}}, \bibinfo {author} {\bibfnamefont {R.}~\bibnamefont
  {{Mandelbaum}}}, \bibinfo {author} {\bibfnamefont {M.}~\bibnamefont
  {{Ishak}}}, \bibinfo {author} {\bibfnamefont {U.}~\bibnamefont {{Seljak}}},
  \bibinfo {author} {\bibfnamefont {R.}~\bibnamefont {{Nichol}}}, \bibinfo
  {author} {\bibfnamefont {K.~A.}\ \bibnamefont {{Pimbblet}}}, \bibinfo
  {author} {\bibfnamefont {N.~P.}\ \bibnamefont {{Ross}}}, \ and\ \bibinfo
  {author} {\bibfnamefont {D.}~\bibnamefont {{Wake}}},\ }\href {\doibase
  10.1111/j.1365-2966.2007.12312.x} {\bibfield  {journal} {\bibinfo  {journal}
  {\mnras}\ }\textbf {\bibinfo {volume} {381}},\ \bibinfo {pages} {1197}
  (\bibinfo {year} {2007})},\ \Eprint {http://arxiv.org/abs/astro-ph/0701671}
  {arXiv:astro-ph/0701671 [astro-ph]} \BibitemShut {NoStop}%
\bibitem [{\citenamefont {{Bridle}}\ and\ \citenamefont
  {{King}}(2007)}]{Bridle}%
  \BibitemOpen
  \bibfield  {author} {\bibinfo {author} {\bibfnamefont {S.}~\bibnamefont
  {{Bridle}}}\ and\ \bibinfo {author} {\bibfnamefont {L.}~\bibnamefont
  {{King}}},\ }\href {\doibase 10.1088/1367-2630/9/12/444} {\bibfield
  {journal} {\bibinfo  {journal} {New Journal of Physics}\ }\textbf {\bibinfo
  {volume} {9}},\ \bibinfo {pages} {444} (\bibinfo {year} {2007})},\ \Eprint
  {http://arxiv.org/abs/0705.0166} {arXiv:0705.0166 [astro-ph]} \BibitemShut
  {NoStop}%
\bibitem [{\citenamefont {{Vlah}}\ \emph {et~al.}(2020)\citenamefont {{Vlah}},
  \citenamefont {{Chisari}},\ and\ \citenamefont {{Schmidt}}}]{Vlah}%
  \BibitemOpen
  \bibfield  {author} {\bibinfo {author} {\bibfnamefont {Z.}~\bibnamefont
  {{Vlah}}}, \bibinfo {author} {\bibfnamefont {N.~E.}\ \bibnamefont
  {{Chisari}}}, \ and\ \bibinfo {author} {\bibfnamefont {F.}~\bibnamefont
  {{Schmidt}}},\ }\href {\doibase 10.1088/1475-7516/2020/01/025} {\bibfield
  {journal} {\bibinfo  {journal} {\jcap}\ }\textbf {\bibinfo {volume} {2020}},\
  \bibinfo {eid} {025} (\bibinfo {year} {2020})},\ \Eprint
  {http://arxiv.org/abs/1910.08085} {arXiv:1910.08085 [astro-ph.CO]}
  \BibitemShut {NoStop}%
\bibitem [{\citenamefont {{Blazek}}\ \emph {et~al.}(2019)\citenamefont
  {{Blazek}}, \citenamefont {{MacCrann}}, \citenamefont {{Troxel}},\ and\
  \citenamefont {{Fang}}}]{Blazekperturb}%
  \BibitemOpen
  \bibfield  {author} {\bibinfo {author} {\bibfnamefont {J.~A.}\ \bibnamefont
  {{Blazek}}}, \bibinfo {author} {\bibfnamefont {N.}~\bibnamefont
  {{MacCrann}}}, \bibinfo {author} {\bibfnamefont {M.~A.}\ \bibnamefont
  {{Troxel}}}, \ and\ \bibinfo {author} {\bibfnamefont {X.}~\bibnamefont
  {{Fang}}},\ }\href {\doibase 10.1103/PhysRevD.100.103506} {\bibfield
  {journal} {\bibinfo  {journal} {\prd}\ }\textbf {\bibinfo {volume} {100}},\
  \bibinfo {eid} {103506} (\bibinfo {year} {2019})},\ \Eprint
  {http://arxiv.org/abs/1708.09247} {arXiv:1708.09247 [astro-ph.CO]}
  \BibitemShut {NoStop}%
\bibitem [{\citenamefont {{Schneider}}\ and\ \citenamefont
  {{Bridle}}(2010)}]{Schneider}%
  \BibitemOpen
  \bibfield  {author} {\bibinfo {author} {\bibfnamefont {M.~D.}\ \bibnamefont
  {{Schneider}}}\ and\ \bibinfo {author} {\bibfnamefont {S.}~\bibnamefont
  {{Bridle}}},\ }\href {\doibase 10.1111/j.1365-2966.2009.15956.x} {\bibfield
  {journal} {\bibinfo  {journal} {\mnras}\ }\textbf {\bibinfo {volume} {402}},\
  \bibinfo {pages} {2127} (\bibinfo {year} {2010})},\ \Eprint
  {http://arxiv.org/abs/0903.3870} {arXiv:0903.3870 [astro-ph.CO]} \BibitemShut
  {NoStop}%
\bibitem [{\citenamefont {{Fortuna}}\ \emph {et~al.}(2020)\citenamefont
  {{Fortuna}} \emph {et~al.}}]{Fortuna}%
  \BibitemOpen
  \bibfield  {author} {\bibinfo {author} {\bibfnamefont {M.~C.}\ \bibnamefont
  {{Fortuna}}} \emph {et~al.},\ }\href@noop {} {\bibfield  {journal} {\bibinfo
  {journal} {arXiv e-prints}\ ,\ \bibinfo {eid} {arXiv:2003.02700}} (\bibinfo
  {year} {2020})},\ \Eprint {http://arxiv.org/abs/2003.02700} {arXiv:2003.02700
  [astro-ph.CO]} \BibitemShut {NoStop}%
\bibitem [{\citenamefont {{Hirata}}\ and\ \citenamefont
  {{Seljak}}(2004)}]{GLHirata}%
  \BibitemOpen
  \bibfield  {author} {\bibinfo {author} {\bibfnamefont {C.~M.}\ \bibnamefont
  {{Hirata}}}\ and\ \bibinfo {author} {\bibfnamefont {U.}~\bibnamefont
  {{Seljak}}},\ }\href {\doibase 10.1103/PhysRevD.70.063526} {\bibfield
  {journal} {\bibinfo  {journal} {\prd}\ }\textbf {\bibinfo {volume} {70}},\
  \bibinfo {eid} {063526} (\bibinfo {year} {2004})},\ \Eprint
  {http://arxiv.org/abs/astro-ph/0406275} {arXiv:astro-ph/0406275 [astro-ph]}
  \BibitemShut {NoStop}%
\bibitem [{\citenamefont {{Chisari}}\ and\ \citenamefont
  {{Dvorkin}}(2013)}]{Chisari}%
  \BibitemOpen
  \bibfield  {author} {\bibinfo {author} {\bibfnamefont {N.~E.}\ \bibnamefont
  {{Chisari}}}\ and\ \bibinfo {author} {\bibfnamefont {C.}~\bibnamefont
  {{Dvorkin}}},\ }\href {\doibase 10.1088/1475-7516/2013/12/029} {\bibfield
  {journal} {\bibinfo  {journal} {\jcap}\ }\textbf {\bibinfo {volume} {2013}},\
  \bibinfo {eid} {029} (\bibinfo {year} {2013})},\ \Eprint
  {http://arxiv.org/abs/1308.5972} {arXiv:1308.5972 [astro-ph.CO]} \BibitemShut
  {NoStop}%
\bibitem [{\citenamefont {{Schmidt}}\ \emph {et~al.}(2015)\citenamefont
  {{Schmidt}}, \citenamefont {{Chisari}},\ and\ \citenamefont
  {{Dvorkin}}}]{Schmidt}%
  \BibitemOpen
  \bibfield  {author} {\bibinfo {author} {\bibfnamefont {F.}~\bibnamefont
  {{Schmidt}}}, \bibinfo {author} {\bibfnamefont {N.~E.}\ \bibnamefont
  {{Chisari}}}, \ and\ \bibinfo {author} {\bibfnamefont {C.}~\bibnamefont
  {{Dvorkin}}},\ }\href {\doibase 10.1088/1475-7516/2015/10/032} {\bibfield
  {journal} {\bibinfo  {journal} {\jcap}\ }\textbf {\bibinfo {volume} {2015}},\
  \bibinfo {eid} {032} (\bibinfo {year} {2015})},\ \Eprint
  {http://arxiv.org/abs/1506.02671} {arXiv:1506.02671 [astro-ph.CO]}
  \BibitemShut {NoStop}%
\bibitem [{\citenamefont {{Okumura}}\ \emph {et~al.}(2019)\citenamefont
  {{Okumura}}, \citenamefont {{Taruya}},\ and\ \citenamefont
  {{Nishimichi}}}]{Okumura1}%
  \BibitemOpen
  \bibfield  {author} {\bibinfo {author} {\bibfnamefont {T.}~\bibnamefont
  {{Okumura}}}, \bibinfo {author} {\bibfnamefont {A.}~\bibnamefont {{Taruya}}},
  \ and\ \bibinfo {author} {\bibfnamefont {T.}~\bibnamefont {{Nishimichi}}},\
  }\href {\doibase 10.1103/PhysRevD.100.103507} {\bibfield  {journal} {\bibinfo
   {journal} {\prd}\ }\textbf {\bibinfo {volume} {100}},\ \bibinfo {eid}
  {103507} (\bibinfo {year} {2019})},\ \Eprint
  {http://arxiv.org/abs/1907.00750} {arXiv:1907.00750 [astro-ph.CO]}
  \BibitemShut {NoStop}%
\bibitem [{\citenamefont {{Meerburg}}\ \emph {et~al.}(2019)\citenamefont
  {{Meerburg}} \emph {et~al.}}]{Meerburg}%
  \BibitemOpen
  \bibfield  {author} {\bibinfo {author} {\bibfnamefont {P.~D.}\ \bibnamefont
  {{Meerburg}}} \emph {et~al.},\ }\href@noop {} {\bibfield  {journal} {\bibinfo
   {journal} {\baas}\ }\textbf {\bibinfo {volume} {51}},\ \bibinfo {eid} {107}
  (\bibinfo {year} {2019})},\ \Eprint {http://arxiv.org/abs/1903.04409}
  {arXiv:1903.04409 [astro-ph.CO]} \BibitemShut {NoStop}%
\bibitem [{\citenamefont {{Okumura}}\ and\ \citenamefont
  {{Taruya}}(2020)}]{OkumuraTaruya}%
  \BibitemOpen
  \bibfield  {author} {\bibinfo {author} {\bibfnamefont {T.}~\bibnamefont
  {{Okumura}}}\ and\ \bibinfo {author} {\bibfnamefont {A.}~\bibnamefont
  {{Taruya}}},\ }\href {\doibase 10.1093/mnrasl/slaa024} {\bibfield  {journal}
  {\bibinfo  {journal} {\mnras}\ }\textbf {\bibinfo {volume} {493}},\ \bibinfo
  {pages} {L124} (\bibinfo {year} {2020})},\ \Eprint
  {http://arxiv.org/abs/1912.04118} {arXiv:1912.04118 [astro-ph.CO]}
  \BibitemShut {NoStop}%
\bibitem [{\citenamefont {{Okumura}}\ \emph {et~al.}(2020)\citenamefont
  {{Okumura}}, \citenamefont {{Taruya}},\ and\ \citenamefont
  {{Nishimichi}}}]{Okumura2}%
  \BibitemOpen
  \bibfield  {author} {\bibinfo {author} {\bibfnamefont {T.}~\bibnamefont
  {{Okumura}}}, \bibinfo {author} {\bibfnamefont {A.}~\bibnamefont {{Taruya}}},
  \ and\ \bibinfo {author} {\bibfnamefont {T.}~\bibnamefont {{Nishimichi}}},\
  }\href {\doibase 10.1093/mnras/staa718} {\bibfield  {journal} {\bibinfo
  {journal} {\mnras}\ }\textbf {\bibinfo {volume} {494}},\ \bibinfo {pages}
  {694} (\bibinfo {year} {2020})},\ \Eprint {http://arxiv.org/abs/2001.05302}
  {arXiv:2001.05302 [astro-ph.CO]} \BibitemShut {NoStop}%
\bibitem [{\citenamefont {{Aghanim}}\ \emph {et~al.}(2018)\citenamefont
  {{Aghanim}} \emph {et~al.}}]{planck}%
  \BibitemOpen
  \bibfield  {author} {\bibinfo {author} {\bibfnamefont {N.}~\bibnamefont
  {{Aghanim}}} \emph {et~al.} (\bibinfo {collaboration} {Planck
  collaboration}),\ }\href@noop {} {\bibfield  {journal} {\bibinfo  {journal}
  {arXiv e-prints}\ ,\ \bibinfo {eid} {arXiv:1807.06209}} (\bibinfo {year}
  {2018})},\ \Eprint {http://arxiv.org/abs/1807.06209} {arXiv:1807.06209
  [astro-ph.CO]} \BibitemShut {NoStop}%
\bibitem [{\citenamefont {{Bernstein}}\ and\ \citenamefont
  {{Jarvis}}(2002)}]{Bernstein}%
  \BibitemOpen
  \bibfield  {author} {\bibinfo {author} {\bibfnamefont {G.~M.}\ \bibnamefont
  {{Bernstein}}}\ and\ \bibinfo {author} {\bibfnamefont {M.}~\bibnamefont
  {{Jarvis}}},\ }\href {\doibase 10.1086/338085} {\bibfield  {journal}
  {\bibinfo  {journal} {\aj}\ }\textbf {\bibinfo {volume} {123}},\ \bibinfo
  {pages} {583} (\bibinfo {year} {2002})},\ \Eprint
  {http://arxiv.org/abs/astro-ph/0107431} {arXiv:astro-ph/0107431 [astro-ph]}
  \BibitemShut {NoStop}%
\bibitem [{\citenamefont {{Blazek}}\ \emph {et~al.}(2011)\citenamefont
  {{Blazek}}, \citenamefont {{McQuinn}},\ and\ \citenamefont
  {{Seljak}}}]{Blazek}%
  \BibitemOpen
  \bibfield  {author} {\bibinfo {author} {\bibfnamefont {J.}~\bibnamefont
  {{Blazek}}}, \bibinfo {author} {\bibfnamefont {M.}~\bibnamefont {{McQuinn}}},
  \ and\ \bibinfo {author} {\bibfnamefont {U.}~\bibnamefont {{Seljak}}},\
  }\href {\doibase 10.1088/1475-7516/2011/05/010} {\bibfield  {journal}
  {\bibinfo  {journal} {\jcap}\ }\textbf {\bibinfo {volume} {2011}},\ \bibinfo
  {eid} {010} (\bibinfo {year} {2011})},\ \Eprint
  {http://arxiv.org/abs/1101.4017} {arXiv:1101.4017 [astro-ph.CO]} \BibitemShut
  {NoStop}%
\bibitem [{\citenamefont {Dodelson}(2003)}]{Dodelson}%
  \BibitemOpen
  \bibfield  {author} {\bibinfo {author} {\bibfnamefont {S.}~\bibnamefont
  {Dodelson}},\ }\href@noop {} {\emph {\bibinfo {title} {Modern cosmology}}},\
  \bibinfo {edition} {1st}\ ed.\ (\bibinfo  {publisher} {Academic Press},\
  \bibinfo {year} {2003})\BibitemShut {NoStop}%
\bibitem [{\citenamefont {{Strauss}}\ and\ \citenamefont
  {{Willick}}(1995)}]{Strauss}%
  \BibitemOpen
  \bibfield  {author} {\bibinfo {author} {\bibfnamefont {M.~A.}\ \bibnamefont
  {{Strauss}}}\ and\ \bibinfo {author} {\bibfnamefont {J.~A.}\ \bibnamefont
  {{Willick}}},\ }\href {\doibase 10.1016/0370-1573(95)00013-7} {\bibfield
  {journal} {\bibinfo  {journal} {\physrep}\ }\textbf {\bibinfo {volume}
  {261}},\ \bibinfo {pages} {271} (\bibinfo {year} {1995})},\ \Eprint
  {http://arxiv.org/abs/astro-ph/9502079} {arXiv:astro-ph/9502079 [astro-ph]}
  \BibitemShut {NoStop}%
\bibitem [{\citenamefont {{Sunyaev}}\ and\ \citenamefont
  {{Zeldovich}}(1980)}]{Sunyaev}%
  \BibitemOpen
  \bibfield  {author} {\bibinfo {author} {\bibfnamefont {R.~A.}\ \bibnamefont
  {{Sunyaev}}}\ and\ \bibinfo {author} {\bibfnamefont {I.~B.}\ \bibnamefont
  {{Zeldovich}}},\ }\href {\doibase 10.1093/mnras/190.3.413} {\bibfield
  {journal} {\bibinfo  {journal} {\mnras}\ }\textbf {\bibinfo {volume} {190}},\
  \bibinfo {pages} {413} (\bibinfo {year} {1980})}\BibitemShut {NoStop}%
\bibitem [{\citenamefont {{Hand}}\ \emph {et~al.}(2012)\citenamefont {{Hand}}
  \emph {et~al.}}]{Hand}%
  \BibitemOpen
  \bibfield  {author} {\bibinfo {author} {\bibfnamefont {N.}~\bibnamefont
  {{Hand}}} \emph {et~al.},\ }\href {\doibase 10.1103/PhysRevLett.109.041101}
  {\bibfield  {journal} {\bibinfo  {journal} {\prl}\ }\textbf {\bibinfo
  {volume} {109}},\ \bibinfo {eid} {041101} (\bibinfo {year} {2012})},\ \Eprint
  {http://arxiv.org/abs/1203.4219} {arXiv:1203.4219 [astro-ph.CO]} \BibitemShut
  {NoStop}%
\bibitem [{\citenamefont {{Smith}}\ \emph {et~al.}(2018)\citenamefont
  {{Smith}}, \citenamefont {{Madhavacheril}}, \citenamefont {{M{\"u}nchmeyer}},
  \citenamefont {{Ferraro}}, \citenamefont {{Giri}},\ and\ \citenamefont
  {{Johnson}}}]{Smith}%
  \BibitemOpen
  \bibfield  {author} {\bibinfo {author} {\bibfnamefont {K.~M.}\ \bibnamefont
  {{Smith}}}, \bibinfo {author} {\bibfnamefont {M.~S.}\ \bibnamefont
  {{Madhavacheril}}}, \bibinfo {author} {\bibfnamefont {M.}~\bibnamefont
  {{M{\"u}nchmeyer}}}, \bibinfo {author} {\bibfnamefont {S.}~\bibnamefont
  {{Ferraro}}}, \bibinfo {author} {\bibfnamefont {U.}~\bibnamefont {{Giri}}}, \
  and\ \bibinfo {author} {\bibfnamefont {M.~C.}\ \bibnamefont {{Johnson}}},\
  }\href@noop {} {\bibfield  {journal} {\bibinfo  {journal} {arXiv e-prints}\
  ,\ \bibinfo {eid} {arXiv:1810.13423}} (\bibinfo {year} {2018})},\ \Eprint
  {http://arxiv.org/abs/1810.13423} {arXiv:1810.13423 [astro-ph.CO]}
  \BibitemShut {NoStop}%
\bibitem [{\citenamefont {{Sugiyama}}\ \emph {et~al.}(2017)\citenamefont
  {{Sugiyama}}, \citenamefont {{Okumura}},\ and\ \citenamefont
  {{Spergel}}}]{Sugiyama}%
  \BibitemOpen
  \bibfield  {author} {\bibinfo {author} {\bibfnamefont {N.~S.}\ \bibnamefont
  {{Sugiyama}}}, \bibinfo {author} {\bibfnamefont {T.}~\bibnamefont
  {{Okumura}}}, \ and\ \bibinfo {author} {\bibfnamefont {D.~N.}\ \bibnamefont
  {{Spergel}}},\ }\href {\doibase 10.1088/1475-7516/2017/01/057} {\bibfield
  {journal} {\bibinfo  {journal} {\jcap}\ }\textbf {\bibinfo {volume} {2017}},\
  \bibinfo {eid} {057} (\bibinfo {year} {2017})},\ \Eprint
  {http://arxiv.org/abs/1606.06367} {arXiv:1606.06367 [astro-ph.CO]}
  \BibitemShut {NoStop}%
\bibitem [{\citenamefont {{Zhu}}\ \emph {et~al.}(2020)\citenamefont {{Zhu}},
  \citenamefont {{White}}, \citenamefont {{Ferraro}},\ and\ \citenamefont
  {{Schaan}}}]{Zhu}%
  \BibitemOpen
  \bibfield  {author} {\bibinfo {author} {\bibfnamefont {H.-M.}\ \bibnamefont
  {{Zhu}}}, \bibinfo {author} {\bibfnamefont {M.}~\bibnamefont {{White}}},
  \bibinfo {author} {\bibfnamefont {S.}~\bibnamefont {{Ferraro}}}, \ and\
  \bibinfo {author} {\bibfnamefont {E.}~\bibnamefont {{Schaan}}},\ }\href
  {\doibase 10.1093/mnras/staa1002} {\bibfield  {journal} {\bibinfo  {journal}
  {\mnras}\ }\textbf {\bibinfo {volume} {494}},\ \bibinfo {pages} {4244}
  (\bibinfo {year} {2020})},\ \Eprint {http://arxiv.org/abs/1910.02318}
  {arXiv:1910.02318 [astro-ph.CO]} \BibitemShut {NoStop}%
\bibitem [{\citenamefont {{Ade}}\ \emph {et~al.}(2019)\citenamefont {{Ade}}
  \emph {et~al.}}]{Simonobs}%
  \BibitemOpen
  \bibfield  {author} {\bibinfo {author} {\bibfnamefont {P.}~\bibnamefont
  {{Ade}}} \emph {et~al.} (\bibinfo {collaboration} {Simons Observatory}),\
  }\href {\doibase 10.1088/1475-7516/2019/02/056} {\bibfield  {journal}
  {\bibinfo  {journal} {\jcap}\ }\textbf {\bibinfo {volume} {2019}},\ \bibinfo
  {eid} {056} (\bibinfo {year} {2019})},\ \Eprint
  {http://arxiv.org/abs/1808.07445} {arXiv:1808.07445 [astro-ph.CO]}
  \BibitemShut {NoStop}%
\bibitem [{\citenamefont {{Tegmark}}\ \emph {et~al.}(1997)\citenamefont
  {{Tegmark}}, \citenamefont {{Taylor}},\ and\ \citenamefont
  {{Heavens}}}]{Tegmark}%
  \BibitemOpen
  \bibfield  {author} {\bibinfo {author} {\bibfnamefont {M.}~\bibnamefont
  {{Tegmark}}}, \bibinfo {author} {\bibfnamefont {A.~N.}\ \bibnamefont
  {{Taylor}}}, \ and\ \bibinfo {author} {\bibfnamefont {A.~F.}\ \bibnamefont
  {{Heavens}}},\ }\href {\doibase 10.1086/303939} {\bibfield  {journal}
  {\bibinfo  {journal} {\apj}\ }\textbf {\bibinfo {volume} {480}},\ \bibinfo
  {pages} {22} (\bibinfo {year} {1997})},\ \Eprint
  {http://arxiv.org/abs/astro-ph/9603021} {arXiv:astro-ph/9603021 [astro-ph]}
  \BibitemShut {NoStop}%
\bibitem [{\citenamefont {{Coe}}(2009)}]{Coe}%
  \BibitemOpen
  \bibfield  {author} {\bibinfo {author} {\bibfnamefont {D.}~\bibnamefont
  {{Coe}}},\ }\href@noop {} {\bibfield  {journal} {\bibinfo  {journal} {arXiv
  e-prints}\ ,\ \bibinfo {eid} {arXiv:0906.4123}} (\bibinfo {year} {2009})},\
  \Eprint {http://arxiv.org/abs/0906.4123} {arXiv:0906.4123 [astro-ph.IM]}
  \BibitemShut {NoStop}%
\bibitem [{\citenamefont {{Taruya}}\ and\ \citenamefont
  {{Okumura}}(2020)}]{Taruya}%
  \BibitemOpen
  \bibfield  {author} {\bibinfo {author} {\bibfnamefont {A.}~\bibnamefont
  {{Taruya}}}\ and\ \bibinfo {author} {\bibfnamefont {T.}~\bibnamefont
  {{Okumura}}},\ }\href {\doibase 10.3847/2041-8213/ab7934} {\bibfield
  {journal} {\bibinfo  {journal} {\apjl}\ }\textbf {\bibinfo {volume} {891}},\
  \bibinfo {eid} {L42} (\bibinfo {year} {2020})},\ \Eprint
  {http://arxiv.org/abs/2001.05962} {arXiv:2001.05962 [astro-ph.CO]}
  \BibitemShut {NoStop}%
\bibitem [{\citenamefont {{Kurita}}\ \emph {et~al.}(2020)\citenamefont
  {{Kurita}}, \citenamefont {{Takada}}, \citenamefont {{Nishimichi}},
  \citenamefont {{Takahashi}}, \citenamefont {{Osato}},\ and\ \citenamefont
  {{Kobayashi}}}]{kurita}%
  \BibitemOpen
  \bibfield  {author} {\bibinfo {author} {\bibfnamefont {T.}~\bibnamefont
  {{Kurita}}}, \bibinfo {author} {\bibfnamefont {M.}~\bibnamefont {{Takada}}},
  \bibinfo {author} {\bibfnamefont {T.}~\bibnamefont {{Nishimichi}}}, \bibinfo
  {author} {\bibfnamefont {R.}~\bibnamefont {{Takahashi}}}, \bibinfo {author}
  {\bibfnamefont {K.}~\bibnamefont {{Osato}}}, \ and\ \bibinfo {author}
  {\bibfnamefont {Y.}~\bibnamefont {{Kobayashi}}},\ }\href@noop {} {\bibfield
  {journal} {\bibinfo  {journal} {arXiv e-prints}\ ,\ \bibinfo {eid}
  {arXiv:2004.12579}} (\bibinfo {year} {2020})},\ \Eprint
  {http://arxiv.org/abs/2004.12579} {arXiv:2004.12579 [astro-ph.CO]}
  \BibitemShut {NoStop}%
\bibitem [{\citenamefont {{Ivezi{\'c}}}\ \emph {et~al.}(2019)\citenamefont
  {{Ivezi{\'c}}} \emph {et~al.}}]{LSST}%
  \BibitemOpen
  \bibfield  {author} {\bibinfo {author} {\bibfnamefont {{\v{Z}}.}~\bibnamefont
  {{Ivezi{\'c}}}} \emph {et~al.} (\bibinfo {collaboration} {LSST}),\ }\href
  {\doibase 10.3847/1538-4357/ab042c} {\bibfield  {journal} {\bibinfo
  {journal} {\apj}\ }\textbf {\bibinfo {volume} {873}},\ \bibinfo {eid} {111}
  (\bibinfo {year} {2019})},\ \Eprint {http://arxiv.org/abs/0805.2366}
  {arXiv:0805.2366 [astro-ph]} \BibitemShut {NoStop}%
\bibitem [{\citenamefont {{Richard}}\ \emph
  {et~al.}(2019{\natexlab{a}})\citenamefont {{Richard}} \emph
  {et~al.}}]{4most}%
  \BibitemOpen
  \bibfield  {author} {\bibinfo {author} {\bibfnamefont {J.}~\bibnamefont
  {{Richard}}} \emph {et~al.} (\bibinfo {collaboration} {CRS Team}),\ }\href
  {\doibase 10.18727/0722-6691/5127} {\bibfield  {journal} {\bibinfo  {journal}
  {The Messenger}\ }\textbf {\bibinfo {volume} {175}},\ \bibinfo {pages} {50}
  (\bibinfo {year} {2019}{\natexlab{a}})},\ \Eprint
  {http://arxiv.org/abs/1903.02474} {arXiv:1903.02474 [astro-ph.CO]}
  \BibitemShut {NoStop}%
\bibitem [{\citenamefont {{Richard}}\ \emph
  {et~al.}(2019{\natexlab{b}})\citenamefont {{Richard}}, \citenamefont
  {{Kneib}}, \citenamefont {{Blake}}, \citenamefont {{Raichoor}}, \citenamefont
  {{Comparat}}, \citenamefont {{Shanks}}, \citenamefont {{Sorce}},
  \citenamefont {{Sahl{\'e}n}}, \citenamefont {{Howlett}}, \citenamefont
  {{Tempel}}, \citenamefont {{McMahon}}, \citenamefont {{Bilicki}},
  \citenamefont {{Roukema}}, \citenamefont {{Loveday}}, \citenamefont
  {{Pryer}}, \citenamefont {{Buchert}}, \citenamefont {{Zhao}},\ and\
  \citenamefont {{CRS Team}}}]{Richard}%
  \BibitemOpen
  \bibfield  {author} {\bibinfo {author} {\bibfnamefont {J.}~\bibnamefont
  {{Richard}}}, \bibinfo {author} {\bibfnamefont {J.~P.}\ \bibnamefont
  {{Kneib}}}, \bibinfo {author} {\bibfnamefont {C.}~\bibnamefont {{Blake}}},
  \bibinfo {author} {\bibfnamefont {A.}~\bibnamefont {{Raichoor}}}, \bibinfo
  {author} {\bibfnamefont {J.}~\bibnamefont {{Comparat}}}, \bibinfo {author}
  {\bibfnamefont {T.}~\bibnamefont {{Shanks}}}, \bibinfo {author}
  {\bibfnamefont {J.}~\bibnamefont {{Sorce}}}, \bibinfo {author} {\bibfnamefont
  {M.}~\bibnamefont {{Sahl{\'e}n}}}, \bibinfo {author} {\bibfnamefont
  {C.}~\bibnamefont {{Howlett}}}, \bibinfo {author} {\bibfnamefont
  {E.}~\bibnamefont {{Tempel}}}, \bibinfo {author} {\bibfnamefont
  {R.}~\bibnamefont {{McMahon}}}, \bibinfo {author} {\bibfnamefont
  {M.}~\bibnamefont {{Bilicki}}}, \bibinfo {author} {\bibfnamefont
  {B.}~\bibnamefont {{Roukema}}}, \bibinfo {author} {\bibfnamefont
  {J.}~\bibnamefont {{Loveday}}}, \bibinfo {author} {\bibfnamefont
  {D.}~\bibnamefont {{Pryer}}}, \bibinfo {author} {\bibfnamefont
  {T.}~\bibnamefont {{Buchert}}}, \bibinfo {author} {\bibfnamefont
  {C.}~\bibnamefont {{Zhao}}}, \ and\ \bibinfo {author} {\bibnamefont {{CRS
  Team}}},\ }\href {\doibase 10.18727/0722-6691/5127} {\bibfield  {journal}
  {\bibinfo  {journal} {The Messenger}\ }\textbf {\bibinfo {volume} {175}},\
  \bibinfo {pages} {50} (\bibinfo {year} {2019}{\natexlab{b}})},\ \Eprint
  {http://arxiv.org/abs/1903.02474} {arXiv:1903.02474 [astro-ph.CO]}
  \BibitemShut {NoStop}%
\bibitem [{\citenamefont {{Chisari}}\ \emph {et~al.}(2019)\citenamefont
  {{Chisari}}, \citenamefont {{Alonso}}, \citenamefont {{Krause}},
  \citenamefont {{Leonard}}, \citenamefont {{Bull}}, \citenamefont {{Neveu}},
  \citenamefont {{Villarreal}}, \citenamefont {{Singh}}, \citenamefont
  {{McClintock}}, \citenamefont {{Ellison}}, \citenamefont {{Du}},
  \citenamefont {{Zuntz}}, \citenamefont {{Mead}}, \citenamefont {{Joudaki}},
  \citenamefont {{Lorenz}}, \citenamefont {{Tr{\"o}ster}}, \citenamefont
  {{Sanchez}}, \citenamefont {{Lanusse}}, \citenamefont {{Ishak}},
  \citenamefont {{Hlozek}}, \citenamefont {{Blazek}}, \citenamefont
  {{Campagne}}, \citenamefont {{Almoubayyed}}, \citenamefont {{Eifler}},
  \citenamefont {{Kirby}}, \citenamefont {{Kirkby}}, \citenamefont
  {{Plaszczynski}}, \citenamefont {{Slosar}}, \citenamefont {{Vrastil}},
  \citenamefont {{Wagoner}},\ and\ \citenamefont {{LSST Dark Energy Science
  Collaboration}}}]{ccl}%
  \BibitemOpen
  \bibfield  {author} {\bibinfo {author} {\bibfnamefont {N.~E.}\ \bibnamefont
  {{Chisari}}}, \bibinfo {author} {\bibfnamefont {D.}~\bibnamefont {{Alonso}}},
  \bibinfo {author} {\bibfnamefont {E.}~\bibnamefont {{Krause}}}, \bibinfo
  {author} {\bibfnamefont {C.~D.}\ \bibnamefont {{Leonard}}}, \bibinfo {author}
  {\bibfnamefont {P.}~\bibnamefont {{Bull}}}, \bibinfo {author} {\bibfnamefont
  {J.}~\bibnamefont {{Neveu}}}, \bibinfo {author} {\bibfnamefont
  {A.}~\bibnamefont {{Villarreal}}}, \bibinfo {author} {\bibfnamefont
  {S.}~\bibnamefont {{Singh}}}, \bibinfo {author} {\bibfnamefont
  {T.}~\bibnamefont {{McClintock}}}, \bibinfo {author} {\bibfnamefont
  {J.}~\bibnamefont {{Ellison}}}, \bibinfo {author} {\bibfnamefont
  {Z.}~\bibnamefont {{Du}}}, \bibinfo {author} {\bibfnamefont {J.}~\bibnamefont
  {{Zuntz}}}, \bibinfo {author} {\bibfnamefont {A.~e.}\ \bibnamefont {{Mead}}},
  \bibinfo {author} {\bibfnamefont {S.}~\bibnamefont {{Joudaki}}}, \bibinfo
  {author} {\bibfnamefont {C.~S.}\ \bibnamefont {{Lorenz}}}, \bibinfo {author}
  {\bibfnamefont {T.}~\bibnamefont {{Tr{\"o}ster}}}, \bibinfo {author}
  {\bibfnamefont {J.}~\bibnamefont {{Sanchez}}}, \bibinfo {author}
  {\bibfnamefont {F.}~\bibnamefont {{Lanusse}}}, \bibinfo {author}
  {\bibfnamefont {M.}~\bibnamefont {{Ishak}}}, \bibinfo {author} {\bibfnamefont
  {R.}~\bibnamefont {{Hlozek}}}, \bibinfo {author} {\bibfnamefont
  {J.}~\bibnamefont {{Blazek}}}, \bibinfo {author} {\bibfnamefont {J.-E.}\
  \bibnamefont {{Campagne}}}, \bibinfo {author} {\bibfnamefont
  {H.}~\bibnamefont {{Almoubayyed}}}, \bibinfo {author} {\bibfnamefont
  {T.}~\bibnamefont {{Eifler}}}, \bibinfo {author} {\bibfnamefont
  {M.}~\bibnamefont {{Kirby}}}, \bibinfo {author} {\bibfnamefont
  {D.}~\bibnamefont {{Kirkby}}}, \bibinfo {author} {\bibfnamefont
  {S.}~\bibnamefont {{Plaszczynski}}}, \bibinfo {author} {\bibfnamefont
  {A.}~\bibnamefont {{Slosar}}}, \bibinfo {author} {\bibfnamefont
  {M.}~\bibnamefont {{Vrastil}}}, \bibinfo {author} {\bibfnamefont {E.~L.}\
  \bibnamefont {{Wagoner}}}, \ and\ \bibinfo {author} {\bibnamefont {{LSST Dark
  Energy Science Collaboration}}},\ }\href {\doibase 10.3847/1538-4365/ab1658}
  {\bibfield  {journal} {\bibinfo  {journal} {\apjs}\ }\textbf {\bibinfo
  {volume} {242}},\ \bibinfo {eid} {2} (\bibinfo {year} {2019})},\ \Eprint
  {http://arxiv.org/abs/1812.05995} {arXiv:1812.05995 [astro-ph.CO]}
  \BibitemShut {NoStop}%
\bibitem [{\citenamefont {{Hirata}}(2009)}]{Hirata}%
  \BibitemOpen
  \bibfield  {author} {\bibinfo {author} {\bibfnamefont {C.~M.}\ \bibnamefont
  {{Hirata}}},\ }\href {\doibase 10.1111/j.1365-2966.2009.15353.x} {\bibfield
  {journal} {\bibinfo  {journal} {\mnras}\ }\textbf {\bibinfo {volume} {399}},\
  \bibinfo {pages} {1074} (\bibinfo {year} {2009})},\ \Eprint
  {http://arxiv.org/abs/0903.4929} {arXiv:0903.4929 [astro-ph.CO]} \BibitemShut
  {NoStop}%
\bibitem [{\citenamefont {{Krause}}\ and\ \citenamefont
  {{Hirata}}(2011)}]{Krause11}%
  \BibitemOpen
  \bibfield  {author} {\bibinfo {author} {\bibfnamefont {E.}~\bibnamefont
  {{Krause}}}\ and\ \bibinfo {author} {\bibfnamefont {C.~M.}\ \bibnamefont
  {{Hirata}}},\ }\href {\doibase 10.1111/j.1365-2966.2010.17638.x} {\bibfield
  {journal} {\bibinfo  {journal} {\mnras}\ }\textbf {\bibinfo {volume} {410}},\
  \bibinfo {pages} {2730} (\bibinfo {year} {2011})},\ \Eprint
  {http://arxiv.org/abs/1004.3611} {arXiv:1004.3611 [astro-ph.CO]} \BibitemShut
  {NoStop}%
\bibitem [{\citenamefont {{Martens}}\ \emph {et~al.}(2018)\citenamefont
  {{Martens}}, \citenamefont {{Hirata}}, \citenamefont {{Ross}},\ and\
  \citenamefont {{Fang}}}]{Martens}%
  \BibitemOpen
  \bibfield  {author} {\bibinfo {author} {\bibfnamefont {D.}~\bibnamefont
  {{Martens}}}, \bibinfo {author} {\bibfnamefont {C.~M.}\ \bibnamefont
  {{Hirata}}}, \bibinfo {author} {\bibfnamefont {A.~J.}\ \bibnamefont
  {{Ross}}}, \ and\ \bibinfo {author} {\bibfnamefont {X.}~\bibnamefont
  {{Fang}}},\ }\href {\doibase 10.1093/mnras/sty1100} {\bibfield  {journal}
  {\bibinfo  {journal} {\mnras}\ }\textbf {\bibinfo {volume} {478}},\ \bibinfo
  {pages} {711} (\bibinfo {year} {2018})},\ \Eprint
  {http://arxiv.org/abs/1802.07708} {arXiv:1802.07708 [astro-ph.CO]}
  \BibitemShut {NoStop}%
\bibitem [{\citenamefont {{Obuljen}}\ \emph {et~al.}(2019)\citenamefont
  {{Obuljen}}, \citenamefont {{Dalal}},\ and\ \citenamefont
  {{Percival}}}]{Obuljen1}%
  \BibitemOpen
  \bibfield  {author} {\bibinfo {author} {\bibfnamefont {A.}~\bibnamefont
  {{Obuljen}}}, \bibinfo {author} {\bibfnamefont {N.}~\bibnamefont {{Dalal}}},
  \ and\ \bibinfo {author} {\bibfnamefont {W.~J.}\ \bibnamefont {{Percival}}},\
  }\href {\doibase 10.1088/1475-7516/2019/10/020} {\bibfield  {journal}
  {\bibinfo  {journal} {\jcap}\ }\textbf {\bibinfo {volume} {2019}},\ \bibinfo
  {eid} {020} (\bibinfo {year} {2019})},\ \Eprint
  {http://arxiv.org/abs/1906.11823} {arXiv:1906.11823 [astro-ph.CO]}
  \BibitemShut {NoStop}%
\bibitem [{\citenamefont {{Obuljen}}\ \emph {et~al.}(2020)\citenamefont
  {{Obuljen}}, \citenamefont {{Percival}},\ and\ \citenamefont
  {{Dalal}}}]{Obuljen2}%
  \BibitemOpen
  \bibfield  {author} {\bibinfo {author} {\bibfnamefont {A.}~\bibnamefont
  {{Obuljen}}}, \bibinfo {author} {\bibfnamefont {W.~J.}\ \bibnamefont
  {{Percival}}}, \ and\ \bibinfo {author} {\bibfnamefont {N.}~\bibnamefont
  {{Dalal}}},\ }\href {\doibase 10.1088/1475-7516/2020/10/058} {\bibfield
  {journal} {\bibinfo  {journal} {\jcap}\ }\textbf {\bibinfo {volume} {2020}},\
  \bibinfo {eid} {058} (\bibinfo {year} {2020})},\ \Eprint
  {http://arxiv.org/abs/2004.07240} {arXiv:2004.07240 [astro-ph.CO]}
  \BibitemShut {NoStop}%
\bibitem [{\citenamefont {{Leonard}}\ \emph {et~al.}(2018)\citenamefont
  {{Leonard}}, \citenamefont {{Mandelbaum}},\ and\ \citenamefont {{LSST Dark
  Energy Science Collaboration}}}]{Leonard}%
  \BibitemOpen
  \bibfield  {author} {\bibinfo {author} {\bibfnamefont {C.~D.}\ \bibnamefont
  {{Leonard}}}, \bibinfo {author} {\bibfnamefont {R.}~\bibnamefont
  {{Mandelbaum}}}, \ and\ \bibinfo {author} {\bibnamefont {{LSST Dark Energy
  Science Collaboration}}},\ }\href {\doibase 10.1093/mnras/sty1444} {\bibfield
   {journal} {\bibinfo  {journal} {\mnras}\ }\textbf {\bibinfo {volume}
  {479}},\ \bibinfo {pages} {1412} (\bibinfo {year} {2018})},\ \Eprint
  {http://arxiv.org/abs/1802.08263} {arXiv:1802.08263 [astro-ph.CO]}
  \BibitemShut {NoStop}%
\bibitem [{\citenamefont {{Singh}}\ and\ \citenamefont
  {{Mandelbaum}}(2016)}]{Singh&Mandelbaum}%
  \BibitemOpen
  \bibfield  {author} {\bibinfo {author} {\bibfnamefont {S.}~\bibnamefont
  {{Singh}}}\ and\ \bibinfo {author} {\bibfnamefont {R.}~\bibnamefont
  {{Mandelbaum}}},\ }\href {\doibase 10.1093/mnras/stw144} {\bibfield
  {journal} {\bibinfo  {journal} {\mnras}\ }\textbf {\bibinfo {volume} {457}},\
  \bibinfo {pages} {2301} (\bibinfo {year} {2016})},\ \Eprint
  {http://arxiv.org/abs/1510.06752} {arXiv:1510.06752 [astro-ph.CO]}
  \BibitemShut {NoStop}%
\bibitem [{\citenamefont {{Georgiou}}\ \emph {et~al.}(2019)\citenamefont
  {{Georgiou}}, \citenamefont {{Chisari}}, \citenamefont {{Fortuna}},
  \citenamefont {{Hoekstra}}, \citenamefont {{Kuijken}}, \citenamefont
  {{Joachimi}}, \citenamefont {{Vakili}}, \citenamefont {{Bilicki}},
  \citenamefont {{Dvornik}}, \citenamefont {{Erben}}, \citenamefont {{Giblin}},
  \citenamefont {{Heymans}}, \citenamefont {{Napolitano}},\ and\ \citenamefont
  {{Shan}}}]{Georgiou}%
  \BibitemOpen
  \bibfield  {author} {\bibinfo {author} {\bibfnamefont {C.}~\bibnamefont
  {{Georgiou}}}, \bibinfo {author} {\bibfnamefont {N.~E.}\ \bibnamefont
  {{Chisari}}}, \bibinfo {author} {\bibfnamefont {M.~C.}\ \bibnamefont
  {{Fortuna}}}, \bibinfo {author} {\bibfnamefont {H.}~\bibnamefont
  {{Hoekstra}}}, \bibinfo {author} {\bibfnamefont {K.}~\bibnamefont
  {{Kuijken}}}, \bibinfo {author} {\bibfnamefont {B.}~\bibnamefont
  {{Joachimi}}}, \bibinfo {author} {\bibfnamefont {M.}~\bibnamefont
  {{Vakili}}}, \bibinfo {author} {\bibfnamefont {M.}~\bibnamefont {{Bilicki}}},
  \bibinfo {author} {\bibfnamefont {A.}~\bibnamefont {{Dvornik}}}, \bibinfo
  {author} {\bibfnamefont {T.}~\bibnamefont {{Erben}}}, \bibinfo {author}
  {\bibfnamefont {B.}~\bibnamefont {{Giblin}}}, \bibinfo {author}
  {\bibfnamefont {C.}~\bibnamefont {{Heymans}}}, \bibinfo {author}
  {\bibfnamefont {N.~R.}\ \bibnamefont {{Napolitano}}}, \ and\ \bibinfo
  {author} {\bibfnamefont {H.}~\bibnamefont {{Shan}}},\ }\href {\doibase
  10.1051/0004-6361/201935810} {\bibfield  {journal} {\bibinfo  {journal}
  {\aap}\ }\textbf {\bibinfo {volume} {628}},\ \bibinfo {eid} {A31} (\bibinfo
  {year} {2019})},\ \Eprint {http://arxiv.org/abs/1905.00370} {arXiv:1905.00370
  [astro-ph.CO]} \BibitemShut {NoStop}%
\bibitem [{\citenamefont {{Tenneti}}\ \emph {et~al.}(2015)\citenamefont
  {{Tenneti}}, \citenamefont {{Mandelbaum}}, \citenamefont {{Di Matteo}},
  \citenamefont {{Kiessling}},\ and\ \citenamefont {{Khandai}}}]{Tenneti}%
  \BibitemOpen
  \bibfield  {author} {\bibinfo {author} {\bibfnamefont {A.}~\bibnamefont
  {{Tenneti}}}, \bibinfo {author} {\bibfnamefont {R.}~\bibnamefont
  {{Mandelbaum}}}, \bibinfo {author} {\bibfnamefont {T.}~\bibnamefont {{Di
  Matteo}}}, \bibinfo {author} {\bibfnamefont {A.}~\bibnamefont {{Kiessling}}},
  \ and\ \bibinfo {author} {\bibfnamefont {N.}~\bibnamefont {{Khandai}}},\
  }\href {\doibase 10.1093/mnras/stv1625} {\bibfield  {journal} {\bibinfo
  {journal} {\mnras}\ }\textbf {\bibinfo {volume} {453}},\ \bibinfo {pages}
  {469} (\bibinfo {year} {2015})},\ \Eprint {http://arxiv.org/abs/1505.03124}
  {arXiv:1505.03124 [astro-ph.CO]} \BibitemShut {NoStop}%
\bibitem [{\citenamefont {{Velliscig}}\ \emph {et~al.}(2015)\citenamefont
  {{Velliscig}} \emph {et~al.}}]{Velliscig}%
  \BibitemOpen
  \bibfield  {author} {\bibinfo {author} {\bibfnamefont {M.}~\bibnamefont
  {{Velliscig}}} \emph {et~al.},\ }\href {\doibase 10.1093/mnras/stv1690}
  {\bibfield  {journal} {\bibinfo  {journal} {\mnras}\ }\textbf {\bibinfo
  {volume} {453}},\ \bibinfo {pages} {721} (\bibinfo {year} {2015})},\ \Eprint
  {http://arxiv.org/abs/1504.04025} {arXiv:1504.04025 [astro-ph.GA]}
  \BibitemShut {NoStop}%
\bibitem [{\citenamefont {{Chisari}}\ \emph {et~al.}(2015)\citenamefont
  {{Chisari}}, \citenamefont {{Codis}}, \citenamefont {{Laigle}}, \citenamefont
  {{Dubois}}, \citenamefont {{Pichon}}, \citenamefont {{Devriendt}},
  \citenamefont {{Slyz}}, \citenamefont {{Miller}}, \citenamefont {{Gavazzi}},\
  and\ \citenamefont {{Benabed}}}]{Chisarisim}%
  \BibitemOpen
  \bibfield  {author} {\bibinfo {author} {\bibfnamefont {N.}~\bibnamefont
  {{Chisari}}}, \bibinfo {author} {\bibfnamefont {S.}~\bibnamefont {{Codis}}},
  \bibinfo {author} {\bibfnamefont {C.}~\bibnamefont {{Laigle}}}, \bibinfo
  {author} {\bibfnamefont {Y.}~\bibnamefont {{Dubois}}}, \bibinfo {author}
  {\bibfnamefont {C.}~\bibnamefont {{Pichon}}}, \bibinfo {author}
  {\bibfnamefont {J.}~\bibnamefont {{Devriendt}}}, \bibinfo {author}
  {\bibfnamefont {A.}~\bibnamefont {{Slyz}}}, \bibinfo {author} {\bibfnamefont
  {L.}~\bibnamefont {{Miller}}}, \bibinfo {author} {\bibfnamefont
  {R.}~\bibnamefont {{Gavazzi}}}, \ and\ \bibinfo {author} {\bibfnamefont
  {K.}~\bibnamefont {{Benabed}}},\ }\href {\doibase 10.1093/mnras/stv2154}
  {\bibfield  {journal} {\bibinfo  {journal} {\mnras}\ }\textbf {\bibinfo
  {volume} {454}},\ \bibinfo {pages} {2736} (\bibinfo {year} {2015})},\ \Eprint
  {http://arxiv.org/abs/1507.07843} {arXiv:1507.07843 [astro-ph.CO]}
  \BibitemShut {NoStop}%
\end{thebibliography}%

\end{document}